\documentclass[aps,prd,reprint,groupedaddress,superscriptaddress,nofootinbib]{revtex4-2}
\usepackage{orcidlink}
\usepackage{slashed}
\usepackage{bbold}
\usepackage{enumitem}
\usepackage{graphicx}
\usepackage[T1]{fontenc}
\usepackage[utf8]{inputenc}
\usepackage[force]{feynmp-auto}
\usepackage{float}
\usepackage{multirow}
\usepackage{subdepth} 
\usepackage[caption=false]{subfig}
\definecolor{debug}{RGB}{0, 0, 0} 
\newcommand{\coloredmath}[2]{%
	\newcommand{#1}{\ensuremath{{\color{debug}{#2}}}}%
}%

\coloredmath{\Rangle}{r}
\coloredmath{\RPhaseInput}{\omega_{22}}
\coloredmath{\RPhaseReplaced}{\omega_{32}}
\coloredmath{\RPhaseOverall}{\phi_R}
\coloredmath{\majoPhase}{\eta_2}

\coloredmath{\br}{\operatorname{BR}}
\coloredmath{\rate}{\operatorname{CR}}
\coloredmath{\reals}{\mathbb{R}}
\coloredmath{\im}{\operatorname{Im}}
\coloredmath{\sign}{\operatorname{sign}}
\coloredmath{\plane}{\RPhaseInput-\Rangle\text{ plane}}
\coloredmath{\factorplane}{\boxFactor-\photonFactor\text{ plane}}
\coloredmath{\positiveSolution}{\RPhaseReplaced_+}
\coloredmath{\negativeSolution}{\RPhaseReplaced_-}

\coloredmath{\amu}{(g-2)_\mu}
\coloredmath{\mueg}{\mu\to e\gamma}
\coloredmath{\taueg}{\tau\to e\gamma}
\coloredmath{\taumug}{\tau\to \mu\gamma}
\coloredmath{\mueee}{\mu\to3e}
\coloredmath{\taueee}{\tau\to 3e}
\coloredmath{\taummm}{\tau\to 3\mu}
\coloredmath{\taumee}{\tau\to \mu ee}
\coloredmath{\tauemm}{\tau\to e\mu\mu}
\coloredmath{\muec}{\mu Al\to eAl}
\coloredmath{\photondominance}{R_{\text{ph. dom.}}}

\coloredmath{\chargedMass}{m_{H^\pm}}
\coloredmath{\photonFactor}{|\Lambda|m_{H^\pm}^2}
\coloredmath{\boxFactor}{\Lambda^2m_{H^\pm}^2}
\coloredmath{\higgsplane}{\photonFactor-\boxFactor\text{ plane}}

\newcommand{\code}[1]{\texttt{#1}}
\newcommand{\indexvalign}{\ensuremath{^{\vphantom{i}}}}

\newcommand{\hc}{\ensuremath{h.c.}}
\newcommand{\pole}[1]{{\color{debug}\ensuremath{m_{#1}^{\text{pole}}}}}
\newcommand{\tree}[1]{{\color{debug}\ensuremath{m_{#1}^{\text{tree}}}}}

\newcommand{\nuratio}{{\color{debug}\ensuremath{t_{32}}}}

\newcommand{\flavorY}[2]{\ensuremath{{\color{debug}Y^{(#1)}_{#2}}}}

\newcommand{\fullRotation}[1]{%
	\ifnum#1=3\ensuremath{{\color{debug}U}}%
	\else\ifnum#1=4\ensuremath{{\color{debug}\tilde{U}}}%
	\fi\fi
}

\newcommand{\niceRotation}[1]{%
	\ifnum#1=3\ensuremath{{\color{debug}V}}%
		\else\ifnum#1=4\ensuremath{{\color{debug}\tilde{V}}}%
	\fi\fi
}

\newcommand{\seesawRotation}[1]{%
	\ifnum#1=3\ensuremath{{\color{debug}S}}%
		\else\ifnum#1=4\ensuremath{{\color{debug}\tilde{S}}}%
	\fi\fi
}

\newcommand{\loopRotation}[1]{%
	\ifnum#1=2\ensuremath{{\color{debug}\hat{R}}}%
		\else\ifnum#1=3\ensuremath{{\color{debug}R}}%
			\else\ifnum#1=4\ensuremath{{\color{debug}\tilde{R}}}%
	\fi\fi\fi
}

\newcommand{\pmns}[1]{{\color{debug}\ensuremath{U^{#1}_{\text{\gls*{pmns}}}}}}

\usepackage{soul}
\usepackage{changes}
\definechangesauthor[name={Vytautas}, color=blue!30!green]{V}
\definechangesauthor[name={Thomas}, color=green!30!gray]{tg}
\definechangesauthor[name={Uladzimir}, color=cyan!80!gray]{uk}
\definechangesauthor[name={Dominik}, color=blue!30!gray]{DS}

\sethighlightmarkup{\color{authorcolor}\uwave{#1}}

\usepackage[acronym, nohypertypes={acronym}]{glossaries-extra}

\setabbreviationstyle[acronym]{long-short}

\glssetcategoryattribute{acronym}{nohyperfirst}{true}

\newacronym{sm}{SM}{Standard model}
\newacronym{gnm}{GNM}{Grimus-Neufeld model}
\newacronym{thdm}{2HDM}{two-Higgs-doublet model}
\newacronym{vev}{VEV}{vacuum expectation value}
\newacronym{pmns}{PMNS}{Pontecorvo-Maki-Nakagawa-Sakata}
\newacronym{no}{NO}{Normal ordering}   
\newacronym{io}{IO}{Inverted ordering} 
\newacronym{gim}{GIM}{Glashow-Iliopoulos-Maiani}
\newacronym{clfv}{cLFV}{Charged Lepton Flavour Violation}

\newacronym{no1}{NO-1}{realistic benchmark point}
\newacronym{no2}{NO-2}{intermediate benchmark point}
\newacronym{no3}{NO-3}{hopeless benchmark point}

\begin{document}

\title{Box enhanced Charged Lepton Flavor Violation in the Grimus-Neufeld model}

\author{Vytautas~D\=ud\.enas\,\orcidlink{0000-0001-9405-9959}}
\email{vytautasdudenas@inbox.lt}
\affiliation{
Institute of Theoretical Physics and Astronomy, Faculty of Physics, Vilnius University,
9 Sault\.ekio, LT-10222 Vilnius, Lithuania%
}

\author{Thomas~Gajdosik\,\orcidlink{0000-0002-4355-8878}}
\email{thomas.gajdosik@ff.vu.lt}

\affiliation{
Institute of Theoretical Physics and Astronomy, Faculty of Physics, Vilnius University,
9 Sault\.ekio, LT-10222 Vilnius, Lithuania%
}

\author{Uladzimir~Khasianevich\,\orcidlink{0000-0003-0255-0674}}
\email{uladzimir.khasianevich@tu-dresden.de}

\affiliation{
Institut f\"ur Kern- und Teilchenphysik,
TU Dresden, Zellescher Weg 19, 01069 Dresden, Germany
}

\author{Wojciech~Kotlarski\,\orcidlink{0000-0002-1191-6343}}
\email{wojciech.kotlarski@ncbj.gov.pl}

\affiliation{
	National Centre for Nuclear Research, Pasteura 7, 02-093 Warsaw, Poland
}

\author{Dominik~St\"ockinger}
\email{dominik.stoeckinger@tu-dresden.de}
\affiliation{
Institut f\"ur Kern- und Teilchenphysik,
TU Dresden, Zellescher Weg 19, 01069 Dresden, Germany
}

\begin{abstract}
In the Grimus-Neufeld model~(GNM) the neutrino mass generation from
	an extended Higgs sector leads to bounds for Charged Lepton Flavour Violating
	(cLFV) processes. Here we update bounds from a previous study by
	extending the parameter space to a nonvanishing
	Majorana phase of the Pontecorvo-Maki-Nakagawa-Sakata matrix
	and to heavier charged Higgs boson masses.
	Three-body cLFV decays are shown to contribute significantly in the large mass regions, as the boxes are enhanced relatively to photonic diagrams. 
	This is in contrast to the smaller mass region studied before,
	in which the two-body decays tightly restrict the parameter space.
	The Majorana phase is shown to change limits within one order of
	magnitude.  
	The tiny seesaw scale is assumed, which makes the cLFV decays in the GNM similar to the cLFV decays in the scotogenic model and the scoto-seesaw models. 
\end{abstract}

\maketitle
\flushbottom

\section{Introduction}\label{sec:intro}
The smallness and hierarchies of neutrino masses might be explained in
models featuring radiative mass generation. Three simple
examples are provided by the scotogenic model \cite{Ma:2006km}, the
scoto-seesaw model \cite{Rojas:2018wym} and
the \gls{gnm} \cite{Grimus:1989pu}. Specifically the \gls{gnm} is an
economical model with only one single sterile neutrino $N$ and the
extended scalar sector of the \gls{thdm}. At tree level, the seesaw mechanism
generates only a single non-vanishing neutrino mass, governed by a
$Z_2$-odd effective Yukawa coupling $y$; at the one-loop level, loops
involving the extra Higgs states generate a second non-vanishing
neutrino mass, governed by the Peccei-Quinn symmetry breaking Higgs potential parameter
$\lambda_5$. An appealing parameter region of the \gls{gnm} studied in
Refs.~\cite{Dudenas:2022qcw, Dudenas:2022von} is the ``tiny'' seesaw scale region,
where the sterile Majorana mass is below the electroweak scale and the
Peccei-Quinn and $Z_2$ breaking parameters are small.

In neutrino mass models such as the \gls{gnm} or the scotogenic or
scoto-seesaw models, the neutrino mass generation automatically also
implies the existence of  \gls{clfv} processes, studied in~\cite{Toma:2013zsa,  Vicente:2014wga,Rojas:2018wym, Mandal:2021yph} for high sterile neutrino masses. 
In fact, in the three
mentioned models, the predictions for \gls{clfv} processes become
similar (or even identical, for the scoto-seesaw and \gls{gnm} models)
in the case of a tiny sterile neutrino mass, as discussed in \cite{Dudenas:2022von}, thus our study also complements the ones in~\cite{Toma:2013zsa,  Vicente:2014wga,Rojas:2018wym, Mandal:2021yph}.

In  Ref.~\cite{Dudenas:2022von}, the interplay between the neutrino
sector, the scalar sector and \gls{clfv} was used to
analyze which restrictions are imposed  by \gls{clfv} on the scalar
sector.  A main finding was that a
single parameter combination, called photon factor, is constrained
by \gls{clfv}, if the charged Higgs mass is sufficiently small (less then 1 TeV).  In most of the parameter space, the decay \mueg\ was
most constraining, but in order to obtain absolute, most conservative
bounds, special parameter regions needed to be considered in
which \taueg\ or \taumug\ were important.

Ref.~\cite{Dudenas:2022von} focused on the tiny seesaw scale region
of the \gls{gnm}, but several additional restrictions on the parameter
regions were imposed in order to simplify the analysis. The mass of the
charged Higgs was constrained to be less
than 1~TeV and the free Majorana phase was set to zero.
In the present
paper we relax those restrictions and complete the phenomenological study
of this \gls{gnm} scenario in the tiny seesaw scale, by studying
the effects of the free Majorana phase and including  large values
of the charged Higgs mass. A major implication of the extended
parameter space is that further decays such as \taueee\ and \taummm\
need to be considered because the related box diagram contributions
can now be significant.

In section \ref{sec:model} we provide a more detailed summary of
the \gls{gnm} and of the main findings of
Ref.~\cite{Dudenas:2022von}. We also provide an overview of the
changes in the considered parameter space and the expected effects.
Section \ref{sec:pheno} is devoted to the extended phenomenological
analysis, and section \ref{sec:conclusions} presents our conclusions.

\section{The model and its parameters}\label{sec:model}
We will briefly describe the main features of the \gls{gnm} to introduce our definitions. 
For a more detailed explanation of the model we refer to \cite{Dudenas:2022von}.

The \gls{gnm} consists of the general \gls{thdm} with an added gauge singlet Weyl fermion $N$ with a Majorana mass $M$, called the sterile neutrino.
This allows for an additional Yukawa coupling with the singlet $N$. 
New terms to the \gls{thdm} Lagrangian read as:
\begin{equation}
\mathcal{L} \ni 
-\frac{1}{2}MNN
-\flavorY{i}{j} \ell_j\indexvalign\epsilon H_i\indexvalign N+\hc
\label{eq:lepton neutrino yukawa}
\end{equation}
The matrix $\epsilon=i\sigma_2$ combines the two doublets to an $SU(2)$ invariant product, $i$ is the Higgs family index, while $j$ is the flavor index. 
We can always use the Higgs basis, where the Yukawa coupling \flavorY{1}{} to the Higgs doublet with non-vanishing \gls{vev} enables the seesaw mechanism, while \flavorY{2}{} leads to  radiative neutrino mass generation for light neutrinos, resulting in two non-vanishing light neutrino masses at one-loop. 

We assume an approximate $Z_2$ symmetry, under which $H_2$ is odd and $H_1$ is even. The Yukawa coupling \flavorY1{} then explicitly breaks this symmetry by a tiny amount. 
To reproduce the light neutrino masses at an eV scale with a seesaw mechanism, one then must have a tiny mass for the sterile neutrino. 
We call this the ``tiny'' seesaw scale and assume:
\begin{equation}
 M \approx m_4  < 10\, \text{GeV} \Leftrightarrow \sum_i |\flavorY{1}{i}|^2 < 10^{-14}\,,
\end{equation} 
where $m_4$ is the pole mass of the sterile neutrino.
This makes \flavorY{1}{i} to be at least an order of magnitude smaller than the electron Yukawa coupling of the SM. 
In the general \gls{thdm} the Yukawa interactions of charged leptons with $H_2$ would also break the $Z_2$ symmetry, thus we set them to zero. 

The parameter that is responsible for radiative mass generation, and
thus relates the scalar and neutrino sectors, can be defined via the neutrino self energy diagram at vanishing
external momentum:  
\begin{equation}
\Lambda := 
\begin{gathered}
\includegraphics[scale = 0.8]{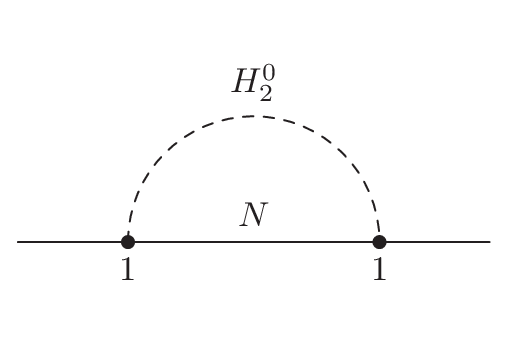}
\end{gathered}\,.\label{eq:Lambda def}
\end{equation} 
In the loop,  $H_2^0$ is the neutral component of the second
Higgs doublet in the Higgs basis and  $N$ is the sterile neutrino. As
indicated, the coupling values have been normalized to unity such that
$\Lambda$ essentially corresponds to the mass-dependent loop function.
For the inert-like scalar sector and the tiny seesaw scenario, this
parameter reduces to 
\begin{equation}
\Lambda = \frac{m_4 }{32 \pi^2} \ln \bigg( \frac{m_H^2}{m^2_A} \bigg)\,,\quad m_4\ll v
\,, 
\label{eq:Lambda for inert}
\end{equation}
where $m_4$ is the mass of the sterile neutrino and $m_{H,A}$ are the
masses of the extra neutral CP-even/odd Higgs bosons. 
We note, however, that our results on \gls{clfv} directly apply for the most general scalar potential, using the definition of Eq.~\eqref{eq:Lambda def}.

In the tiny seesaw limit we are able to get the one-loop parameterization of Yukawa couplings \flavorY{i}{}, which automatically reproduce the observed \gls{pmns} matrix and the neutrino mass differences squared~\cite{Dudenas:2022von}. 	
For \gls{clfv} rates, the Yukawa coupling of neutrinos to the first Higgs doublet in the Higgs basis can be neglected \cite{Dudenas:2022von} and thus only the Yukawa coupling to the second Higgs doublet is important for us, which is: 
\begin{align}
\flavorY{2}{}  &\equiv
\sign(\Lambda)\,
\sqrt{\frac{\pole{2}}{|z\Lambda|}}\,
\left[\begin{matrix}
	0 
	\\
	\cos\Rangle~e^{i\RPhaseInput}
	\\
	\sin\Rangle~e^{i\RPhaseReplaced}
\end{matrix}\right]^T
U
\,,
\label{eq:Y2 mass eigenstates+}
\\
\RPhaseReplaced{} &\equiv
-\frac{1}{2}\arcsin\left(\frac{\sin(2\RPhaseInput{})}{\nuratio \tan^{2}\Rangle}\right)
\,,\quad \nuratio\equiv \frac{\pole{3}}{\pole{2}}\,,
\label{eq:positive sol}
\\
z 
&= 
z(\Rangle\,, \RPhaseInput\,) 
\equiv \cos^2 \Rangle\ e^{2i \RPhaseInput} +
\nuratio  \sin^2 \Rangle\ e^{2i \RPhaseReplaced}
\,,
\label{eq:def z}
\end{align}
where the neutrino masses \pole{2}, \pole{3}, and the mixing matrix $U$, have different values for \gls{no} and \gls{io} and are related to the experimental values \cite{Zyla:2020zbs} as in table~\ref{tab:to experiment},
where
the matrix $O_{\text{\gls*{io}}}$ relates our mixing matrix convention to the usual $3\nu$ convention, used in Ref.~\cite{Zyla:2020zbs}, in the \gls{io} case:
\begin{equation}
O_{\text{\gls*{io}}}=\bigg(\begin{matrix}
0 & 1\\
\mathbb{1}_{2\times 2} & 0
\end{matrix}\bigg)
\,.
\end{equation}

The \gls{pmns} matrix is defined as \citep{Zyla:2020zbs}:
\begin{widetext}
\begin{equation}
\pmns{} =\left(\begin{matrix}1 & 0 & 0\\
0 & c_{23} & s_{23}\\
0 & -s_{23} & c_{23}
\end{matrix}\right)\left(\begin{matrix}c_{13} & 0 & s_{13}e^{-i\delta_{\text{CP}}}\\
0 & 1 & 0\\
-s_{13}e^{i\delta_{\text{CP}}} & 0 & c_{13}
\end{matrix}\right)\left(\begin{matrix}c_{12} & s_{12} & 0\\
-s_{12} & c_{12} & 0\\
0 & 0 & 1
\end{matrix}\right)\left(\begin{matrix}e^{i\eta_{1}} & 0 & 0\\
0 & e^{i\majoPhase{}} & 0\\
0 & 0 & 1
\end{matrix}\right),\label{eq:pmns}
\end{equation}
\end{widetext}
where $\eta_{1}$ and \majoPhase{} are unknown Majorana phases, and $s_{ij}=\sin\theta_{ij}$, and $c_{ij}=\cos\theta_{ij}$. 

\begin{table}
	\centering
	\begin{tabular}{| c | c | c | c| }
		\hline
		& 
		\pole{2} & \pole{3} & $U$
		\\
		\hline	
		\rule{0pt}{3.5ex} 
		\gls*{no} & $\sqrt{\Delta m_{21}^{2}}$ & $\sqrt{|\Delta m_{32}^{2}|+\Delta m_{21}^{2}}$ & 
		\pmns{\dagger}
		\\
		\hline	
		\rule{0pt}{3.5ex} 
		\gls*{io} & $\sqrt{|\Delta m_{32}^{2}|- \Delta m_{21}^{2} }$ & $\sqrt{|\Delta m_{32}^{2}| }$ & $O_{\text{\gls*{io}}}^{\vphantom{\dagger}}\pmns{\dagger} $
		\\
		\hline	
	\end{tabular}
	\caption{Relations of neutrino pole masses and mixings, to the oscillations parameters for \gls{no} and \gls{io}.}
	\label{tab:to experiment}
\end{table}

While the parameters \Rangle{} and \RPhaseInput{} are free, not all values reproduce neutrino masses and mixings. 
They are restricted by $z\in \mathbb{R}$ (see \cite{Dudenas:2022von}). 
$\Lambda$ enters as a free parameter in Eq.~\eqref{eq:Y2 mass eigenstates+}. 
Every other parameter is to be determined by the neutrino masses and mixings. 
However, the Majorana phases $\eta_1$ and \majoPhase{} are not observed so far and thus are, in principle, free parameters. 
In the \gls{gnm}, $\eta_1$ is absorbed into the field redefinition 
and has no physical significance. 
This is the direct consequence of the lightest neutrino being massless in the \gls{gnm}:
a massless neutrino simply does not have a Majorana phase, and hence $\eta_1=0$.
The Majorana phase \majoPhase{}, however, is a free parameter in the \gls{gnm} that does have a physical significance. 
Thus the free parameters that parameterize \flavorY{2}{} are: 
\begin{equation}
\Rangle\,,\quad	\RPhaseReplaced\,, \quad \Lambda \,, \quad \majoPhase{} \,. \label{eq:free in yukawa}
\end{equation}
Note that to simplify the study in \cite{Dudenas:2022von}, we set $\majoPhase{}=0$. 
We now relax this assumption and study its significance for the bounds drawn from \gls{clfv} decays. 

The Yukawa coupling \flavorY{2}{} leads to one-loop generated \gls{clfv} decays with a charged Higgs boson and sterile neutrino in the loop. 
These amplitudes for $\ell_i\to \ell_j \gamma$ and $\ell_i\to \ell_j \ell_k \ell_k$ consist of penguin and box contributions, which are~\cite{Dudenas:2022von}:
\begin{align}
&A_{\text{penguin}}\sim \frac{\flavorY2{i}^* \flavorY2{j}}{m^2_{H^{\pm}} } \sim \frac{1}{\photonFactor}\,,
\label{eq:clfv amplitude1} \\
&A_{\text{box}}\sim \frac{\flavorY2{i}^* \flavorY2{j} |\flavorY2{k}|^2}{m^2_{H^{\pm}}} \sim \frac{1}{\boxFactor}\,, 
\label{eq:clfv amplitudes}
\end{align}
where we used the parameterization from Eq.~\eqref{eq:Y2 mass eigenstates+} to factor out the so-called \emph{photon factor} \photonFactor{} and the \emph{box factor} \boxFactor{}.

It is then  useful to translate the set of free parameters of Eq.~\eqref{eq:free in yukawa}, together with the free mass of the charged Higgs \chargedMass{} into the new set of parameters
\begin{equation}
\Rangle\,,\quad	\RPhaseReplaced\,, \quad \majoPhase{}\,, \quad \photonFactor \,, \quad \boxFactor
\label{eq:free params}
\end{equation}
that directly controls the \gls{clfv} rates. 
 
In the tiny-seesaw parameter region, the \gls{clfv} decays thus dominantly depend on five parameters, Eq.~\eqref{eq:free params}.
Two of them, \photonFactor{} with \boxFactor{}, are connected to the Higgs sector while others parameterize the Yukawa sector only. 

\subsection{Recap of the previous study}

In \cite{Dudenas:2022von} we assumed a relatively low charged Higgs mass
of $\chargedMass{}\lesssim 1$~TeV and $\majoPhase{}=0$. This led to a
suppression of box diagrams, which are essentially governed by the
factor \boxFactor{}, called box factor. 
In addition, three-body decays turned out to be unimportant in that
parameter region. Instead, two-body decays gave the strongest constraints. 
As a result, only three out of the five parameters in Eq.\ (\ref{eq:free
params}) were important, and we could put an experimental bound on the photon
factor \photonFactor{} as a function of the \plane{}
from two-body decays only. 
The used branching ratios are listed in table~\ref{tab:observables-experiments}.
They are categorized by \emph{phases} of experiments: 
the current experimental limits are refered as \emph{current phase}, 
while \emph{next phase} are the planned sensitivities of the upgraded experiments.

{
\setlength\tabcolsep{2pt} 
\begin{table}
	\centering
	\begin{tabular}{| c | c | c | c |}
		\hline
		Process & 
		Current phase & 
		Next phase\\
		\hline	
		\rule{0pt}{3ex} 
		\mueg &
		MEG~\cite{MEG:2020zxk}: $4.2\cdot 10^{-13}$
		&
		MEG-II~\cite{MEGII:2018kmf}: $6\cdot 10^{-14}$
		\\
		\taueg & 
		BaBar~\cite{BaBar:2009hkt}: $3.3\cdot10^{-8}$
		&
		Belle-II~\cite{Belle-II:2018jsg}: $3.0\cdot10^{-9}$
		\\
		\taumug &
		BaBar~\cite{BaBar:2009hkt}:
		$4.5\cdot10^{-8}$
		&
		Belle-II~\cite{Belle-II:2018jsg}: $1.0\cdot10^{-9}$
		\\
		\mueee
		& 
		SINDRUM~\cite{SINDRUM:1987nra}: $1\cdot 10^{-12}$
		&
		Mu3e-I~\cite{Wasili:2020ksf}: $2\cdot 10^{-15}$
		\\
		\taueee &
		Belle-I~\cite{Hayasaka:2010np}: $2.7\cdot10^{-8}$
		&
		Belle-II~\cite{Belle-II:2018jsg}: $4.6\cdot10^{-10}$
		\\
		\taumee &
		Belle-I~\cite{Hayasaka:2010np}: $1.8\cdot10^{-8}$
		&
		Belle-II~\cite{Belle-II:2018jsg}: $3.1\cdot10^{-10}$
		\\
		\tauemm&
		Belle-I~\cite{Hayasaka:2010np}: $2.7\cdot10^{-8}$
		&
		Belle-II~\cite{Belle-II:2018jsg}: $4.6\cdot10^{-10}$
		\\
		\taummm& 
		Belle-I~\cite{Hayasaka:2010np}: $2.1\cdot10^{-8}$
		&
		Belle-II~\cite{Belle-II:2018jsg}: $3.6\cdot10^{-10}$ 
		\\
		\hline
	\end{tabular}
	\caption{Current and next experimental bounds, divided into different phases, related to corresponding observables. Data for $\tau$ decays for Belle-II was obtained from the figure~189 of Ref.~\cite{Belle-II:2018jsg}.} 
	\label{tab:observables-experiments}	
\end{table}
}

The most constraining experiment for the photon factor in almost all the \plane{} is \mueg{}.
However, there exist fine-tuned areas, in which this decay rate vanishes, because of the vanishing of either \flavorY2{e} or \flavorY2{\mu}. 
The point in the \plane{} where the corresponding Yukawa
coupling \flavorY2{e,\mu} vanishes (and thus, the corresponding decay rate) one gets \cite{Dudenas:2022von} from Eq.~\eqref{eq:Y2 mass eigenstates+}:
\begin{equation}
\cot( \Rangle ) e^{i (\RPhaseInput - \RPhaseReplaced) }= - \nuratio \frac{U_{3 f}}{U_{2 f}}
\label{eq:Y2 zero condition}
\end{equation}
for flavor $f=e,\mu$.
There is one solution of Eq.~\eqref{eq:Y2 zero condition} for each flavour.
This means that there are only two special regions in which \mueg{}
and \mueee{} decays are suppressed: around the points in the  \plane{}  with~$\flavorY2{e,\mu}=0$.  

In these special regions, the $\tau$ decay experiments give stronger constraints than $\mu$ decays. 
The experimental sensitivities of $\tau$ decays
(table~\ref{tab:observables-experiments}) give much weaker constraints on the photon factor. 
This also means that in most of the \plane{}, i.e.\ outside the
special regions, an observation of $\tau$ decays in the experiments with the planned sensitivities of next phase experiments (second row of table~\ref{tab:observables-experiments}) is already excluded by \mueg{} in the current phase. 

Thus, to specify the regions in a robust way, we define \emph{special}
regions as the regions in the \plane{} where the two-body decays of $\tau$ lepton are
potentially observable at the Belle-II experiment. In short, the
regions are defined by the following inequalities for the theory
predictions for $\mu$ and $\tau$ decays:
\begin{equation}\label{eq:fine-tuned-definition}
\begin{alignedat}{3}
\emph{special region 1:}\quad& &&\\
\br(\text{Belle-II}) <\:& \br(\taueg) &< \br(\text{BaBar})
\\
\text{while}~& \br(\mueg) &< \br(\text{MEG})\,,
\\
\emph{special region 2:}\quad & && \\
 \br(\text{Belle-II}) <\:& \br(\taumug) &< \br(\text{BaBar})
\\
\text{while}~& \br(\mueg) &< \br(\text{MEG})
\,,
\end{alignedat} 
\end{equation}
where by $\br(\text{experiment})$ we indicate the expected (for Belle-II) or current (MEG and BaBar) limits on the branching ratio of a corresponding decay in the experiment.
They occupy relatively small areas in the \plane{}, as one sees in figure~7 of~\cite{Dudenas:2022von} and thus these areas can be dubbed as ``unnatural''/fine-tuned. 

The \emph{absolute} lower bound on the photon factor is defined as the
lowest photon factor value possible anywhere in the \plane{}, for which all the experimental constraints, shown in table~\ref{tab:observables-experiments} are satisfied. 
Naturally, this absolute bound is defined by the limits on the branching ratios of $\tau$ decays in the special regions, as they are the weakest possible constraints. 
However, since the special regions might be seen as unnatural, we also define a \emph{typical} bound on the photon factor, which is the bound got in the same way, but excluding the special regions and thus derived from \mueg{} branching ratio.  

From a practical point of view, the typical bound should coincide with the bound from the rough random scan over the \plane{}. 
In contrast, to derive the absolute bound, one has to use the analytic solution of \eqref{eq:Y2 zero condition}, to get the point in the \plane{} of \mueg{} close to zero to a high degree of accuracy. 
As a result, the typical bound is $O(10^2)$ stronger than the absolute one. 

\subsection{Extending the parameter space}\label{sec:extending par}

\begin{table}
	\centering
	\begin{tabular}{| c | c | c |}
		\hline
		Parameters studied in \cite{Dudenas:2022von}  & 
		Current parameters \\
		\hline	
		$\majoPhase{} = 0$ &
		$-\pi <\majoPhase{} < \pi$ 
		\\
		\hline	
		$ \chargedMass{} < 1$ TeV
		 &
		$ \chargedMass{} < 5$ TeV 
		\\[0.5ex]
		\hline	
		$ |z| >0.5$  & 
		$|z| >0.38$ 
		\\
		\hline	
		$|\flavorY2{i}|<1 $  & 
        $\sum_i |\flavorY2{i}|^2< 8\pi $
		\\[0.5ex]
		\hline
	\end{tabular}
	\caption{Extensions of parameter ranges, compared to the values studied in \cite{Dudenas:2022von}. \label{tab:extension of params}}
\end{table}

The extension of the parameter space in the current study vs. the
study of \cite{Dudenas:2022von} is summarized in
table~\ref{tab:extension of params}.  
Here we briefly discuss the expected effects and related subtleties. 

The extension of $\majoPhase{}$ is rather straightforward 
and allows to study the $\majoPhase{}$  dependence of the limits. 
We expect the  qualitative behavior of the model to remain unchanged
and expect the resulting limits to stay within the same order of
magnitude. For the \gls{io}, however, it turns out that some values of $\majoPhase{}$ push the special regions into the area that we previously excluded in the study due to a too large tree-loop cancellation in the neutrino pole mass calculation. 
This cancellation is encoded in the parameter $z$ of Eq.~\eqref{eq:def z}, which can be shown to be~\cite{Dudenas:2022von}:
\begin{equation}
|z|=\frac{\pole{3}}{\tree{3}}\,,
\end{equation} 
where \tree{3} is the tree-level neutrino mass. 
Hence we decided to extend the considered values of $z$ to allow roughly 62\% cancellation instead of 50\% to incorporate the special regions for all \majoPhase{}. 
 
In the parameter regions where \chargedMass{} is large,
larger Yukawa couplings are allowed. As a result, box diagram
contributions governed by four powers of Yukawa couplings can become relevant.
To see this, consider a fixed \photonFactor{} (which is fixed by the limits of the two-body decay experiments) and put it into Eqs.~(\ref{eq:clfv amplitude1},~\ref{eq:clfv amplitudes}) to verify the following proportionality of the relative importance of box diagrams:
\begin{equation}
A_{\text{box}}/A_{\text{penguin}}
\sim 
1/\Lambda
\stackrel{\photonFactor=\text{const}}{\sim}
m^2_{H^\pm}
\,. \label{eq:box/penguin}
\end{equation}

Hence extending the mass range for \chargedMass{} also increases the
importance of  three-body decays. 
At some very large \chargedMass{}, however,
limits drawn from perturbative unitarity will become more important
than the limits drawn from \gls{clfv} decays. 
  
The restriction of Yukawa coupling values in
Ref.~\cite{Dudenas:2022von} was technically motivated by the
numerical stability of the neutrino pole mass calculation
in \code{FlexibleSUSY}~\cite{Athron:2014yba,Athron:2017fvs, Athron:2021kve}, which also served as a rigorous numerical
check of the parameterization of Eq.~\eqref{eq:Y2 mass eigenstates+}. 
This restriction, even though somewhat ad hoc, was of little consequence
since it was in general weaker than the ones derived from \gls{clfv} in low \chargedMass{} region, thus it did not affect our results drawn from two-body decays. 
We now look at larger \chargedMass{}, drop the ad hoc restriction on the Yukawa coupling, and only
apply the looser restriction derived from perturbative unitarity (see
section~\ref{sec:yukawa-bounds}).
For $\chargedMass{}<1$~TeV the only consequence, compared to our previous limits in Ref. \cite{Dudenas:2022von}, is the change of the absolute bound for \gls{io}, see Eq.~\eqref{eq:absolute}.

We note, however, that for such large Yukawas, the pole mass calculation becomes numerically unstable in \code{FlexibleSUSY} and yields errors in the one-loop output for the sectors that are irrelevant for \gls{clfv}.	
However, the \gls{clfv} rates depend only on the Yukawa couplings and the $\overline{\text{MS}}$ charged Higgs mass \chargedMass{}. 
The parameterization of Yukawa couplings of Eq.~\eqref{eq:Y2 mass eigenstates+} is numerically stable and consistent with one-loop neutrino masses and mixings by construction, as was confirmed numerically. 
The \gls{clfv} observables are independent of the pole mass calculations up to the phase-space integration factor in \code{FlexibleSUSY}.
This allows to speed up parameter scans by selecting the pole mass loop order setting to 0 in \code{FlexibleSUSY} LesHouches input. 
Note, that this setting choice prints the pole masses and mixings inconsistent with the experimental data in the \code{FlexibleSUSY} output and hence they should be ignored for mentioned input option.
We use the \code{FlexibleSUSY} extension \code{NPointFunctions}~\cite{Khasianevich:2022ess} to get three-body decay rates. 

\subsection{Perturbative unitarity bound}\label{sec:yukawa-bounds}

In the low scalar mass regime that was studied
in~\cite{Dudenas:2022von}, the bound on the Yukawa coupling that we
got from two-body \gls{clfv} processes is strong enough, such that
there was no need to care about bounds coming from perturbative
unitarity.  
If the scalar mass \chargedMass{} is  larger,  larger Yukawa couplings
are in agreement with \gls{clfv} bounds and hence motivated. 
Since we are now covering the  large scalar mass regions, we have to be more careful in order to be consistent with the perturbative unitarity constraints for Yukawa couplings. 
Using the bounds described in \cite{Allwicher:2021rtd} and applying them to the \gls{gnm},
we get the requirement: 

\begin{equation}
\textstyle{\sum_i}|\flavorY2{i}|^2 < 8 \pi\,. \label{eq:pert unitarity}
\end{equation}

Using Eq.~\eqref{eq:Y2 mass eigenstates+}, we can translate the
unitarity bound Eq.~\eqref{eq:pert unitarity} into a bound on
$\Lambda$, which depends on $z,\RPhaseInput, \Rangle$:

\begin{equation}
\begin{gathered}
\Lambda > \frac{\pole{2}  f_z\left(\Rangle , \RPhaseInput \right) }{8 \pi}
\,,\\ 
f_z\left(\Rangle , \RPhaseInput \right)=   \frac{\cos^2\Rangle +  \nuratio^2 \sin^2 \Rangle}{z\left(\Rangle, \RPhaseInput \right) } 
\,.
\label{eq:pert bound on Lam}
\end{gathered}
\end{equation}

For $|z|>0.5$, we get the following ranges for $\Lambda$:

\begin{equation}
\begin{aligned}
\text{\gls*{no}:}\quad & \Lambda > (0.34 - 6.1) \cdot 10^{-12}\, \text{GeV}\,,   \\ 
\text{\gls*{io}:}\quad & \Lambda > (1.2 - 4.6) \cdot 10^{-12}\, \text{GeV}\,.
\end{aligned}
\end{equation}

For $\majoPhase{}=0$ at two special points of interest, the perturbative unitarity bounds give:

\begin{equation}
\begin{aligned}
\text{\gls*{no}: } \quad  & 
 \Lambda> 3.7 \cdot 10^{-12\,}\text{GeV for } \flavorY2{e}=0\,,\\
 & \Lambda> 5.0 \cdot 10^{-13}\,\text{GeV for } \flavorY2{\mu}=0\,, \\ 
\text{\gls*{io}: } \quad & \Lambda> 2.2 \cdot 10^{-12}\,\text{GeV for } \flavorY2{e}=0\,, \\
& \Lambda> 2.1 \cdot 10^{-12}\,\text{GeV for } \flavorY2{\mu}=0\,.  \label{eq:pert lambda values}
\end{aligned}
\end{equation}

In \cite{Dudenas:2022von}, we got the absolute bound on the photon factor \photonFactor{} from two-body decays of $O(10^{-6})~\text{GeV}^3$ while the typical one is of $O(10^{-4})~\text{GeV}^3$. 
Comparing it with Eq.~\eqref{eq:pert bound on Lam}, we see that the absolute bound starts to compete with the perturbative unitarity bound for $\chargedMass{}>1$~TeV while we typically should expect stronger bounds from \gls{clfv} up to $\chargedMass{}\sim 10$~TeV. 
Note that such a large value for the Higgs mass here  is problematic from the consistency of the scalar sector alone, hence it is safe to say, that typically we have stronger constraints from \gls{clfv}, period. 
For the previously defined special regions, however, the bound \eqref{eq:pert unitarity} becomes important for  $\chargedMass{}>1$~TeV and thus needs to be taken into account.  

\section{Phenomenological analysis}\label{sec:pheno}
\subsection{The dependence of absolute and typical bounds on Majorana phase (two-body decays)}

We begin our phenomenological analysis by studying the impact of
the Majorana phase $\majoPhase{}$. It influences the value of the
Yukawa couplings in the \plane{}. In particular, the points in the
\plane{} where  $\flavorY2{e,\mu}=0$ vary as a function of
$\majoPhase{}$ as can be obtained analytically from
Eq.~\eqref{eq:Y2 zero condition}.

\begin{widetext}

\begin{figure}
	\hspace*{\fill}
	\subfloat[]{
		\includegraphics[width= 0.4 \textwidth]{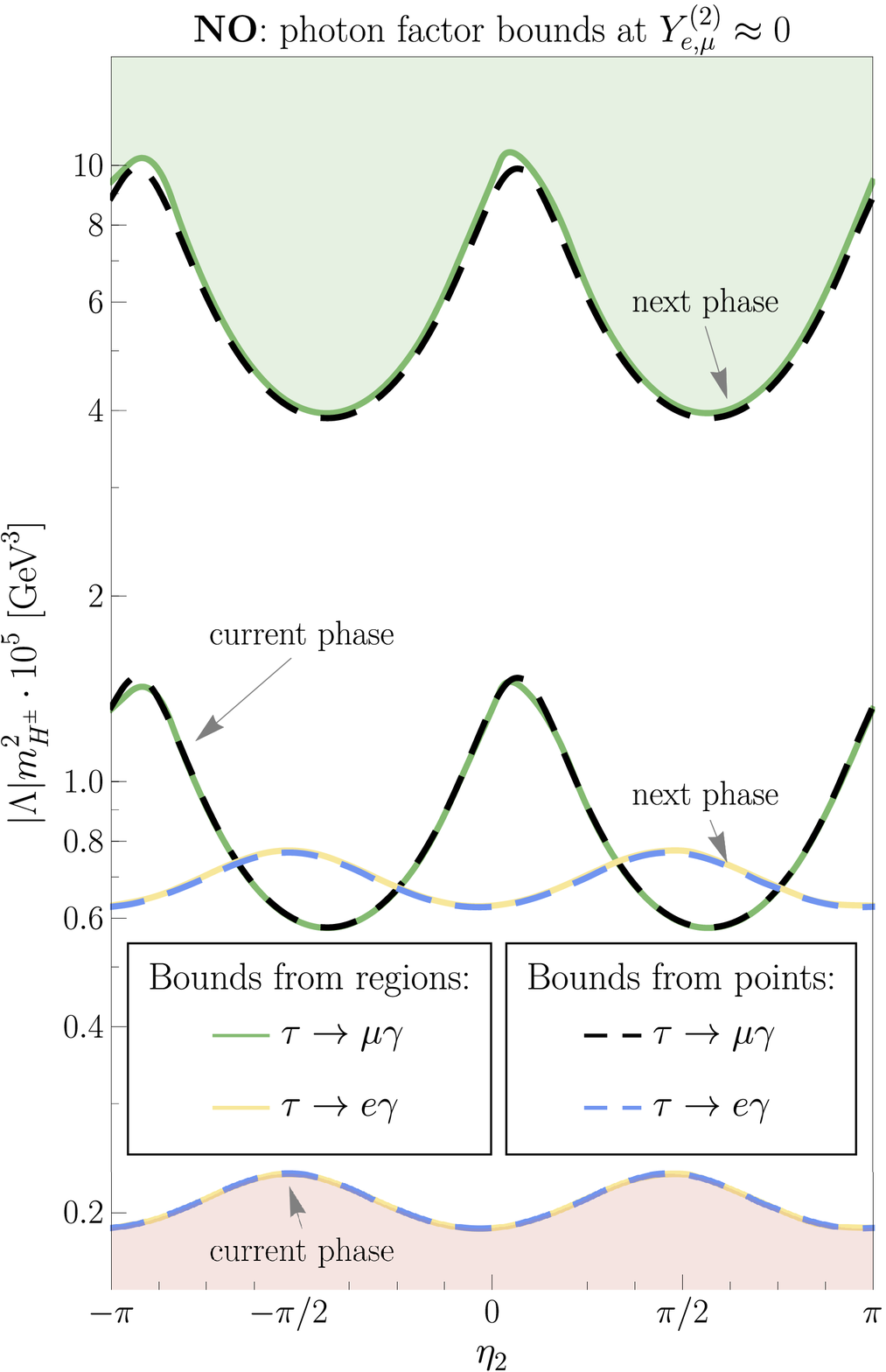}
		\label{fig:no-two-body-bounds} }
	\hfill
	\subfloat[]{
		\includegraphics[width= 0.4 \textwidth]{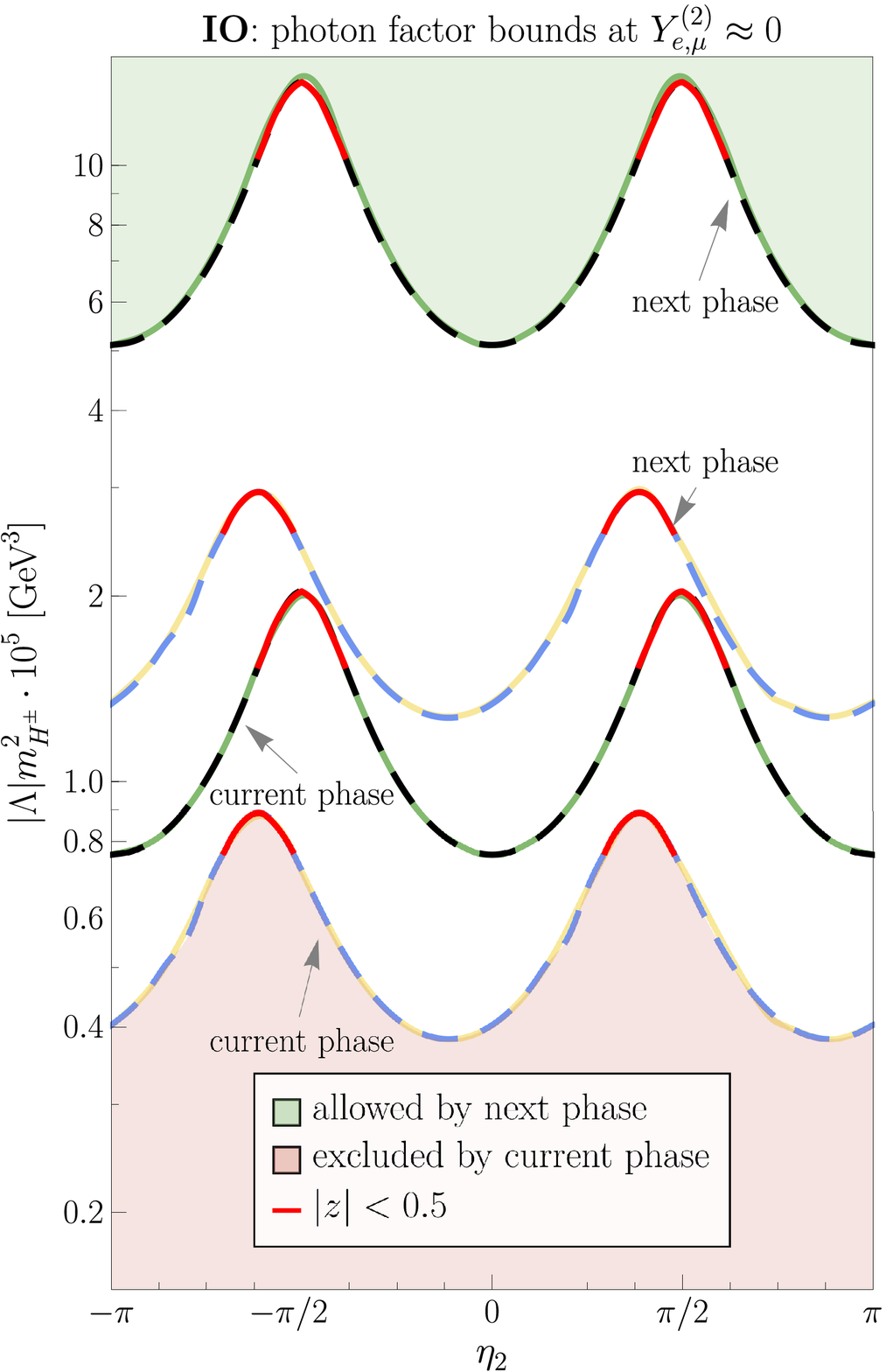}
		\label{fig:io-two-body-bounds} }
	\hspace*{\fill}
	\caption[]{	Bounds on the photon factor \photonFactor{} for two-body $\tau$ decays that come from the 
		special points (extension of table 2 of Ref.~\cite{Dudenas:2022von}) and regions (extension of table 3 of Ref.~\cite{Dudenas:2022von}) for different Majorana phase \majoPhase{} values.
		The highest line corresponds to the special region and was defined as the 
		\emph{typical} limit. The lowest line comes from special points and corresponds to the weakest constraint of the \gls{gnm} and was called \emph{absolute} limit.
		Photon factors below the lines are excluded by the appropriate phase of the two-body decay experiments. 
		This splits the whole area into three subregions: the red area is already excluded by current experiments (current phase from table~\ref{tab:observables-experiments}), the white area that can potentially be excluded by two-body decays of the planned experiments (next phase from table~\ref{tab:observables-experiments}) and the green region that will remain allowed by the limits on the two-body decay in the planned experiments.
		For \gls{io} special points can move into the parameter region in which $|z|<0.5$ for specified ranges of \majoPhase{} (highlighted by the red color). There, $|z|$ varies in the following regions: for $\flavorY{2}{e}=0$ by $|z|\in(0.38, 0.48)$ and for $\flavorY{2}{\mu}=0$ by $|z|\in(0.43, 0.49)$.
	}\label{fig:new-two-body-absolute-limits}
\end{figure}
\end{widetext}

As discussed before, in most of the parameter space the decay
\mueg\ provides the strongest constraint; however the most
conservative (``absolute'') bound is obtained by considering the
two special points where $\flavorY2{e,\mu}=0$ and where \mueg\ becomes
insensitive.
Hence we focus here on the bounds derived from these special points
and study their  $\majoPhase{}$ dependence. The results are shown in
figure~\ref{fig:new-two-body-absolute-limits}.

The blue dashed lines in figure~\ref{fig:new-two-body-absolute-limits}
correspond to the special point where $\flavorY2{\mu}=0$ and which is
only constrained by \taueg. For this special point, all values of the
photon factor below the blue dashed line are excluded by \taueg{} (the lower/upper blue dashed
line corresponds to current/next phase as defined in
table~\ref{tab:observables-experiments}). The yellow line
corresponds to the small region with $\flavorY2{\mu}\approx0$ and
whose boundary is defined by
Eq.\ (\ref{eq:fine-tuned-definition}). Again, the photon factor values
below the line are excluded by \taueg\ in all of this region.

The red shaded area in figure~\ref{fig:new-two-body-absolute-limits}
highlights all photon factor values below the lowest yellow
line. This area is excluded already by the weakest possible
constraint from the current phase experiments in all of the parameter space \textemdash{} hence it is absolutely
excluded already at present. If the next phase of the experiments do not see
a signal, this absolutely excluded area will move up to the upper
yellow line.

Similarly, the black dashed lines correspond to the special point
where  $\flavorY2{e}=0$ and which is
only constrained by \taumug. The green dashed lines correspond to the small
region around this point, defined by
Eq.\ (\ref{eq:fine-tuned-definition}). At this point/in this small
region, all photon factor values below the respective lines are
excluded by \taumug\ but not by other observables
(the lower lines correspond to current phase experiments, the upper
lines to next phase experiments). The limit in this region is stronger than the
absolute bound derived from the yellow dashed line.

The region above the upper green line cannot be excluded by $\tau$
decays even at next phase of the experiments, given existing
\mueg\ constraints, however typically values of the photon factor
close to the green/blue dashed lines are excluded by \mueg{} \textemdash{} hence we
refer to these lines as ``typical'' bounds.
We refer to Ref.~\cite{Dudenas:2022von} for  extensive
discussions of absolute and typical bounds 
and for details on the \mueg\ constraints on
the entire parameter space.

Bounds on the photon factor derived from points $\flavorY2{e,\mu}=0$ and from small regions $\flavorY2{e,\mu}\approx0$, defined by Eq.~\eqref{eq:fine-tuned-definition} are almost the same, as seen in figure~\ref{fig:new-two-body-absolute-limits}.
The small difference in these bounds reflects the smallness of these regions.
Thus from now on, we approximate the bounds that come from special regions by the ones from special points.

Our main result here is the impact of the Majorana phase.
Figure~\ref{fig:new-two-body-absolute-limits} shows that the
dependence on \majoPhase{} value does not lead to a drastic change in
the two-body decay bounds. The variation of the absolute bound  is
either roughly a factor of 1.25 for \gls{no} or a factor of 2 for
\gls{io}. 
The absolute bound for $\majoPhase{}=0$ is actually very close to the lowest
possible value.  Hence, the result for the absolute bound obtained in
\cite{Dudenas:2022von} remains indeed also the absolute lower bound if we include
\majoPhase{} in the analysis as well.

\begin{widetext}

\begin{figure}
	\centering
	\begin{tabular}[b]{c}
		\includegraphics[width= 0.8 \textwidth]{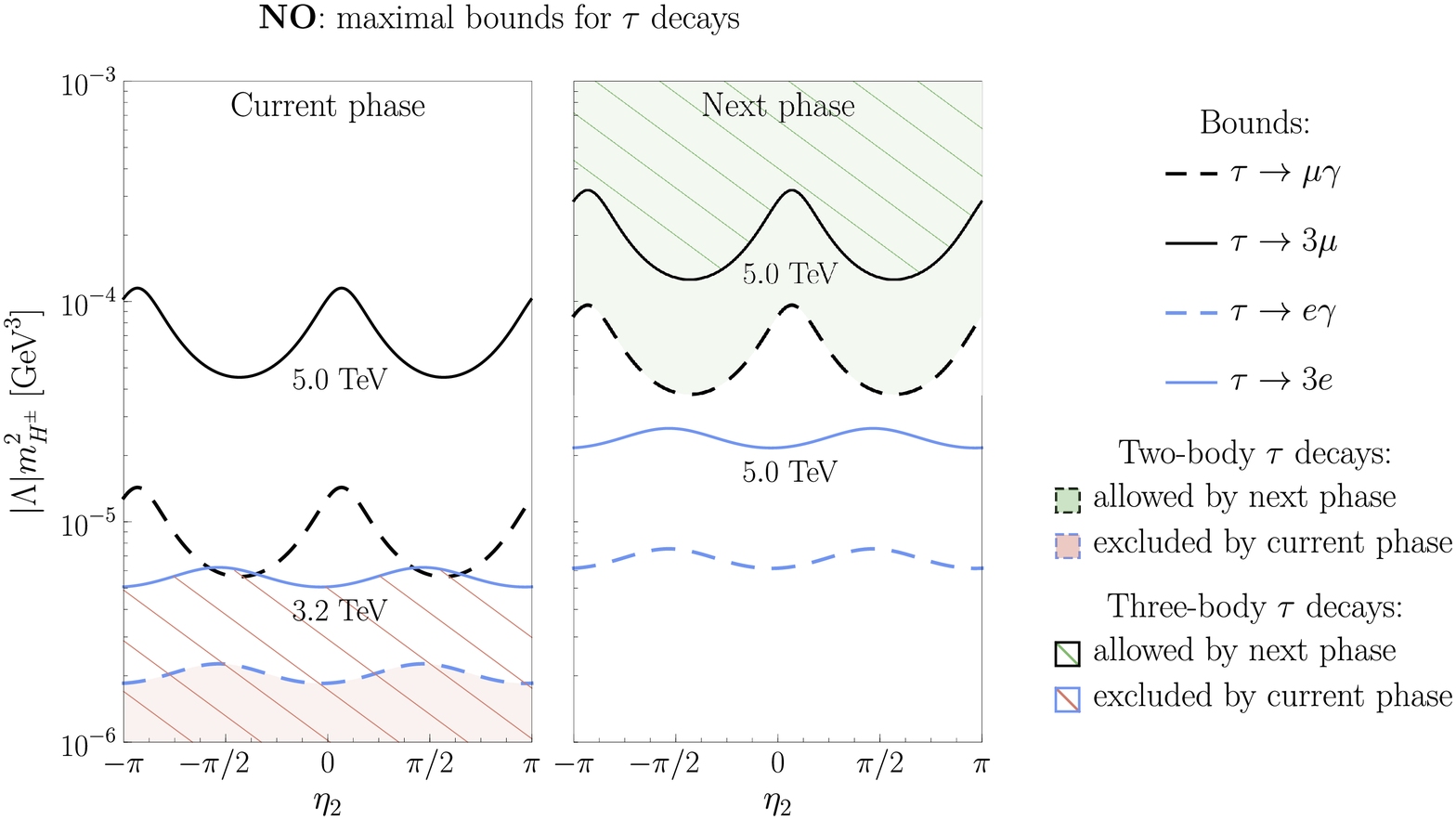} \\
		\includegraphics[width= 0.8 \textwidth]{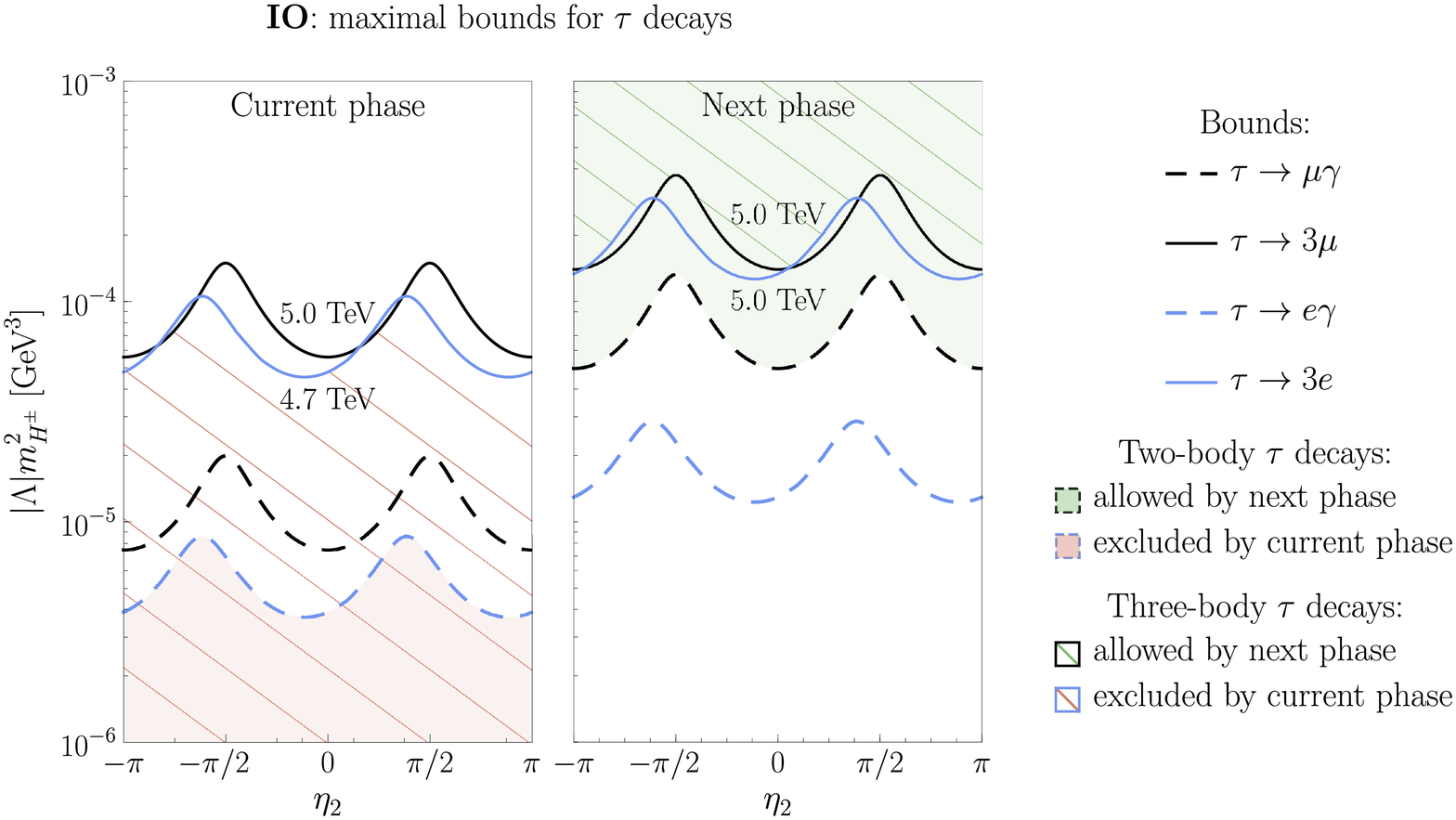}
	\end{tabular}
	\caption[]{ 
		Maximal bounds on the photon factor \photonFactor{} for three-body $\tau$ decays that come from the 	special points for different Majorana phase \majoPhase{} values.
		Two-body decay bounds are shown for the comparison. In general, the limits defined by three-body decays depend on the value of \chargedMass{}. Here, the highest possible lines for three-body decays are shown accompanied with the corresponding charged Higgs boson mass.
		The regions with $|z|<0.5$ exist for three-body decays and are the same as in figure~\ref{fig:new-two-body-absolute-limits}, however they are not shown here in order not to overcrowd the plot.
	}\label{fig:three-body-bounds}
\end{figure}

\end{widetext}

\subsection{The importance of three-body decays}\label{sec:new-three-bounds}

The three-body decays are expected to contribute significantly in regions
of a large Higgs mass. There, for a fixed photon factor value, the box
factor \boxFactor{} becomes smaller (equivalently, $\Lambda$ becomes
smaller and Yukawa couplings are enhanced), so that boxes are enhanced
relatively to photonic diagrams. 
This enhancement might improve the limits that were obtained from two-body decays alone. 
To illustrate this, we consider the same parameter scenarios as in
figure~\ref{fig:new-two-body-absolute-limits} and look at maximum
possible restrictions from the 
three-body decays, to see how
figure~\ref{fig:new-two-body-absolute-limits} is modified.  
The results are shown in figure~\ref{fig:three-body-bounds}, where, 
in order not to overcrowd the pictures, we split bounds from current and next phases of experiments into separate subfigures. 

The dashed lines are the same as in  figure~\ref{fig:new-two-body-absolute-limits}. They correspond to the
constraints from \taueg\ (for the specific point with
$\flavorY2{\mu}=0$) and from \taumug\ (for the specific point with
$\flavorY2{e}=0$). The new solid lines correspond to the
additional limits from three-body decays (\taueee\ for $\flavorY2{\mu}=0$
and \taummm\ for $\flavorY2{e}=0$). These limits depend on
\chargedMass{}; we have plotted the limits for specific values of
\chargedMass{} as indicated. These values are the maximum possible
values for which the bounds, derived from the \gls{clfv} decays are still stronger than the perturbative unitarity bound. 

From figure~\ref{fig:three-body-bounds} we see that three-body
decays can indeed lead to more severe bounds on \photonFactor{}. In fact, the absolute bound for
large \chargedMass{} is now defined by \taueee{}. 
The green (where $\tau$ decays are not expected at Belle-II) and the red (absolutely excluded) shaded regions are now also modified, as shown by hatching.

This highlights that the box amplitudes can change the picture significantly. 
However, we also see that the functional dependence on \majoPhase{} for $\ell \to \ell^\prime \gamma $ and the corresponding $\ell \to 3 \ell^\prime $ are the same. 
Having that in mind, we will now set $\majoPhase{}=0$ again, and study the competition between box and photonic contributions in more detail.

\subsection{Absolute bounds from photon and box factors}\label{sec:23-interplay}

\begin{widetext}
	
\begin{figure}[]
		\hspace*{\fill}
		\subfloat[\label{fig:no-factor-bounds} ]{
			\centering
			\includegraphics[width= 0.45 \textwidth]{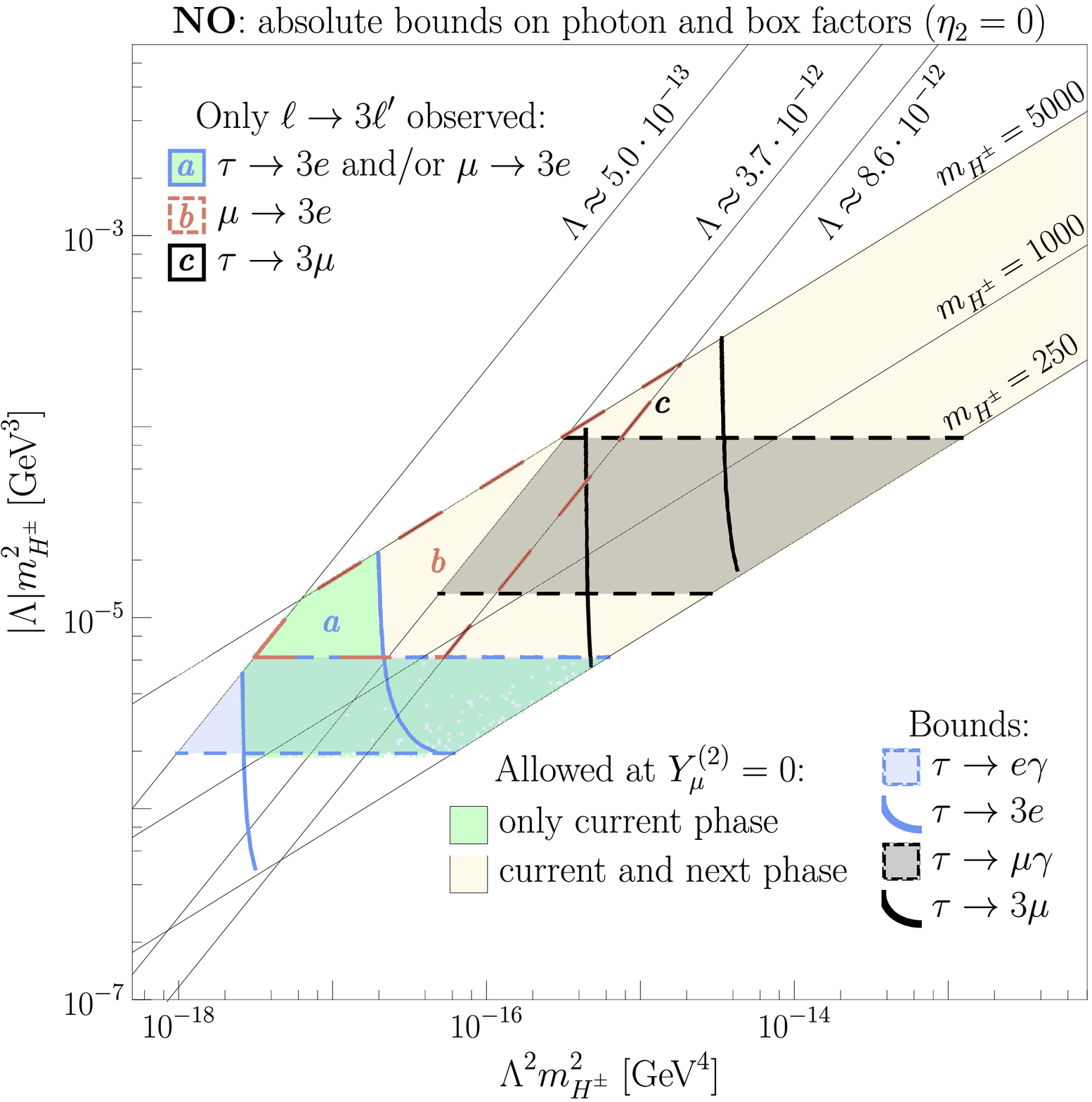} }
		\hfill
		\subfloat[]{
			\centering
			\includegraphics[width= 0.45 \textwidth]{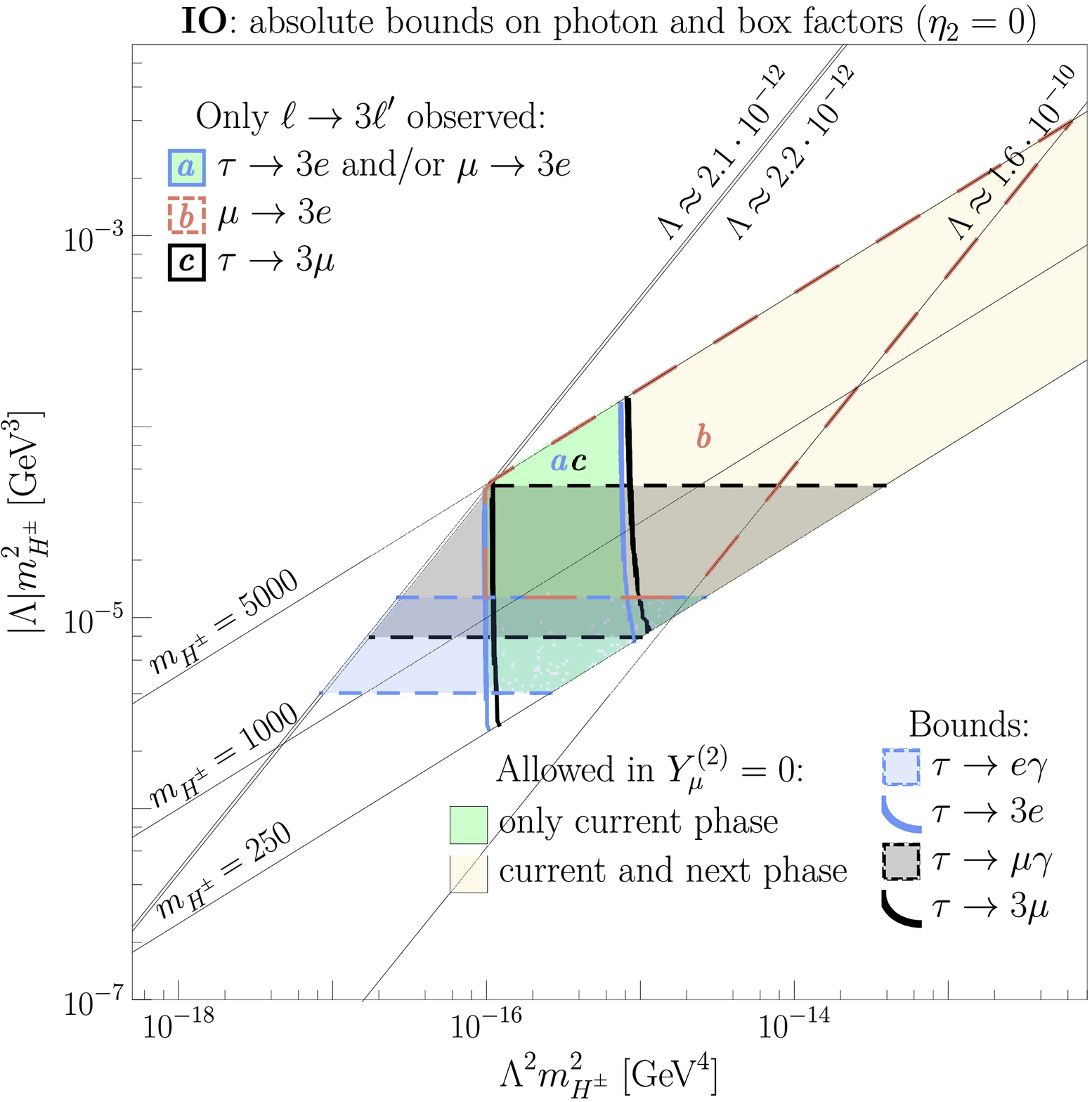} }
		\hspace*{\fill}
		\caption{Constraints from \gls{clfv} shown in the \factorplane{} for $\majoPhase{}=0$.
			Blue/black dashed and solid lines show the constraints of \photonFactor{} and \boxFactor{} from current phase and next phase experiments at the parameter points $\flavorY2{\mu}=0$/$\flavorY2{e}=0$.
			The areas, named by letters, are the areas in which the three-body decays can be observed in the next phase experiments.  
			The rightmost $\Lambda$ values in both plots correspond to the limiting value for which the \mueee{} in Mu3e-I can still be observed. 
			The other shown $\Lambda$ values are the same as in Eq.~\eqref{eq:pert lambda values} and thus give perturbative unitarity constraints at each of the special points (lower $\Lambda$ are excluded). 
		}\label{fig:factor-bounds}
\end{figure}

\end{widetext}

Ref.~\cite{Dudenas:2022von} has shown that the \gls{clfv} amplitudes
in the tiny seesaw parameter region are essentially governed by two
quantities \textemdash{} the box factor 
\boxFactor{} (which was neglected in small \chargedMass{} region) and the photon factor \photonFactor{}. The previous
subsection has shown that the box contributions and three-body decays  can
be significant in the extended parameter region. Here we study the
interplay of all these contributions in more detail.
We study the constraints in the \factorplane{} to see which contributions, 
which processes, or which constraints are more important in
different parameter regions.
We thus set $\majoPhase{}=0$ and plot the constraints from both, two
and three-body decays in the \factorplane{} in
figure~\ref{fig:factor-bounds}.  The plots in
figure~\ref{fig:factor-bounds} contain a wealth of information which
we explain step by step.

First, consider again the bounds from \taueg\ for the specific point
where $\flavorY2{\mu}=0$.  
The two-body decays do not depend on the box factor, thus the current/next
phase constraints from \taueg{} are seen as the horizontal blue
dashed lines, as in figures~\ref{fig:new-two-body-absolute-limits}
and~\ref{fig:three-body-bounds}.  Similarly, the bounds from
\taumug{} for the point where $\flavorY2{e}=0$ are shown by the
horizontal black dashed  lines.  
The shaded areas between these dashed lines correspond to the white
area in figure~\ref{fig:new-two-body-absolute-limits} at
$\majoPhase{}=0$. The absolutely excluded region for two-body decays
(also shown as red area in
figure~\ref{fig:new-two-body-absolute-limits}) is below the
lowest blue dashed line.  

To understand the three body decays, consider, 
for example, the rightmost \taueee{} blue solid line for which the decay rate is the same as the sensitivity in Belle-II. 
Going down this line from highest point to the lowest point corresponds to reducing the value of \chargedMass{}. 
Going down, around $\chargedMass{}\approx400$ GeV, the curve starts to deviate more significantly from the vertical line and close to  $\chargedMass{}=250$ GeV the curve becomes horizontal. 
This means that around those masses, the box contributions become negligible and photonic diagrams  dominate. 
One can see that all the constraints from \taueee{} and \taummm{},
wherever they are allowed by the two-body decays (higher than the corresponding dashed lines) are almost vertical.
This indicates the box dominance of the three-body decays. 

For \gls{no} the region between the blue lines (for \taueg{},
\taueee{} and the point with $\flavorY2{\mu}=0$) does not overlap with
the region between the black lines (for  \taumug{}, \taummm{} and the
point with $\flavorY2{e}=0$). For \gls{io} the corresponding region in
the \factorplane{} overlap significantly.  This is in line with
figure \ref{fig:new-two-body-absolute-limits} (at
$\majoPhase{}=0$). This figure 
also shows that the relative overlap between these regions would
change for other values of $\majoPhase{}$.

The perturbative unitarity limit on $\Lambda$ of Eq.~\eqref{eq:pert
  bound on Lam} can be applied to the special points where $\flavorY2{\mu,e}=0$.
This results in the left borders of the blue/black shaded regions
given by the contour lines of the indicated values of $\Lambda$.
The existence of these borders means e.g.\ that the unitarity limit
becomes stronger than the one coming from \taueee{} in certain
parameter regions. With the help of the contour lines for fixed mass
\chargedMass{}, one can see that for the current experimental limits
(current phase) this happens around $\chargedMass{}\approx 3.2(4.7)$~TeV,
which is consistent with the value shown in
figure~\ref{fig:three-body-bounds}.  
In other cases, next phase experiments put stronger bounds than the
perturbative unitarity limit, as the heaviest mass of the scanned
values is reached. 

To exemplify and highlight the interplay between two- and three-body
decay bounds, we focus again on the special point with
$\flavorY{2}{\mu}=0$, which leads to the most conservative, absolute
bounds valid in all of the parameter space.
The green and yellow shadings in figure~\ref{fig:factor-bounds}
represent the allowed regions at $\flavorY{2}{\mu}=0$. The green
region is allowed by current phase experiments but will be excluded by next phase experiments; the  yellow regions will remain allowed if next phase
experiments do not find a signal. For small charged Higgs masses, the
regions are bounded from below by the horizontal blue dashed lines corresponding
to the two-body decay $\taueg$. If the
mass is increased, then at some value the bound from three-body decays
(\taueee{} in this case, solid blue line) becomes more restrictive. At even higher mass
values, the perturbative unitarity can imply another type of
restriction. 
However, only in the green region for current phase experiments this
restriction is stronger than the one from the experimental
constraints. 
The next phase experimental sensitivities give a stronger bound than the
perturbative unitarity bound in all of the yellow region.

From the green and yellow areas in figure~\ref{fig:factor-bounds} one
can infer at which \chargedMass{} value the three-body decays
become more important than the corresponding two-body ones.  This also
allows to update and generalize the limits  from \gls{clfv} obtained
in \cite{Dudenas:2022von}. 
The most conservative,  ``absolute'' limits,   at the point with $\flavorY2{\mu}=0$  are (for \gls{no} and \gls{io}):
\begin{equation}
\begin{alignedat}{3}
&\photonFactor{} >\,&& 1.9(4.0)\cdot 10^{-6}\,\text{GeV}^3
\\  
&&&\text{with}~\chargedMass{} < 1.2 (0.4) \,\text{TeV}
\,,\\
&\boxFactor{} >\,&& 2.6 (98.7)\cdot 10^{-18} \,\text{GeV}^4
\\
&&&\text{with}~\chargedMass{} \in 1.2\div 3.2 (0.4\div 4.7)\,\text{TeV}\,,
\\
&|\Lambda| >\,&& 0.5(2.1) \cdot 10^{-12} \,\text{GeV}
\\
&&&\text{with}~\chargedMass{} > 3.2(4.7) \,\text{TeV}\,. \label{eq:absolute}
\end{alignedat}
\end{equation}
The first line of Eq.~\eqref{eq:absolute} corresponds to \taueg{}, the second line to \taueee{} and the last to the perturbative unitarity bound. 
The first line of Eq.~\eqref{eq:absolute} also corresponds to the limits in Ref.~\cite{Dudenas:2022von}. For \gls{io} the limit is modified. 
This comes from allowing larger Yukawa couplings and hence excluding lower Higgs masses, when we want to exclude all the Yukawa couplings, as discussed in section~\ref{sec:extending par}.

The region bordered by the upper black (solid and dashed) lines
corresponds to the maximum region where \gls{clfv} $\tau$ decays might
be observed at the next phase experiments. Above that region a large
part of parameter space for generic values of Yukawa couplings can
still be excluded by \gls{clfv} muon decays; hence these boundaries
are referred to as ``typical limits''. These typical limits are not
affected by unitarity, and reading off from the \taumug{} and
\taummm{} next phase experiment lines we obtain for \gls{no} (and \gls{io}):
\begin{equation}
\begin{alignedat}{3}
&\photonFactor{} >\,&& 8.8(4.9)\cdot 10^{-5}\,\text{GeV}^3
\\
&&&\text{with}~\chargedMass{} < 1.5 (1.7) \,\text{TeV}
\,,\\
&\boxFactor{} >\, && 3.4(0.85) \cdot 10^{-15} \,\text{GeV}^4
\\  
&&&\text{with}~\chargedMass{} > 1.5(1.7) \,\text{TeV}\,.\label{eq:typical}
\end{alignedat}
\end{equation}
These limits update the typical limits found in \cite{Dudenas:2022von}. 

Finally, figure~\ref{fig:factor-bounds} also shows the parameter areas
where an observation of \emph{only} three-body decays is possible by next
phase experiments while two-body decays are not observed. We discuss them only
for the left panel of figure~\ref{fig:no-factor-bounds} for \gls{no},
where they are more 
straightforward to localize. The area denoted with the label ``a'' is
above the upper dashed blue line, i.e.\ \taueg{} cannot be
observed at next phase experiments. However, it is to the left of the
solid blue line still, hence \taueee{} is still allowed. Hence this
``a'' area corresponds indeed to the case where 
\taueee{} is observed and \taueg{} is not in forthcoming
experiments. Similarly, the area marked by ``c'' lies above the upper
black dashed line and to the left of the solid black line. In this area
\taummm{} can be observed in the next phase
experiments but \taumug{} cannot. Also, near to
$\flavorY{2}{\mu}\approx 0$ (for both orderings), as was mentioned in
the Ref.~\cite{Dudenas:2022von}, there is a slight deviation from
dipole dominance for \mueee{}. Together with the ratio between MEG-II
and Mu3e-I experimental bounds, this allows to observe \mueee{} while
\mueg{} is not seen. The allowed parameter region is marked ``b'' and
is bordered by the red dashed line. It is also possible to observe
\mueee{} and \taueee{} in the same time due to the aforementioned
localization in the \plane{}, while \mueee{} and \taummm{} cannot be
detected at the same time in the \gls{gnm} for the tiny seesaw
regime. 
The drastic improvement of the sensitivity in Mu3e-II will make
\mueee{} observation possible in all the yellow region of
figure~\ref{fig:factor-bounds} with photon dominated contributions
for both orderings. 

\subsection{Global box dominance of \mueee{} ($\majoPhase{} = 0$)}

Now we focus our attention on the more generic parameter space outside
the special parameter points where $\flavorY2{\mu,e}=0$. Here the
limits from \gls{clfv} muon decays are stronger and $\tau$ decays
cannot be observed at next phase experiments.
In this parameter region the interplay between the two-body decay
\mueg{} and the three-body decay \mueee{} is of interest.
We abbreviate: 
\begin{equation}
R \equiv \frac{\br(\mueee) } {\br(\mueg)}\,.
\end{equation}
For the low Higgs masses studied in \cite{Dudenas:2022qcw}, the
relation between these two observables is simply fixed by the photon
dominance:%
\footnote{%
In the published version of~\cite{Dudenas:2022von}, there is a missing
overall factor of $1/\pi$ in Eq.~(5.1) for the expression of the
photon dominance, see also e.g. Eq.~(29) of \cite{Toma:2013zsa} for
the case of dipole dominance. This misprint does not affect any of the results presented in~\cite{Dudenas:2022von} or here.}
\begin{equation}
\begin{alignedat}{4}
R \approx &\photondominance{} \equiv 
-\frac{5 \alpha}{18 \pi} +\frac{\alpha }{3 \pi } \bigg( \ln\frac{m^2_e}{m^2_\mu} -\frac{11}{4} \bigg) \approx  0.0059
\\ &\text{ for } \chargedMass{} < 1 \text{ TeV} \,. 
\end{alignedat}
\label{eq:photon dominance}
\end{equation}
Note, that the branching ratios for three-body decays
are dominated by photonic contributions in the case of light \chargedMass{}, while others \textemdash{} boxes, $Z$ and Higgs penguins \textemdash{} are negligible.
This regime is called \emph{photon} dominance as it is different from dipole dominance by additional non-negligible vector photon amplitudes, see~\cite{Dudenas:2022von}.
However, this is no longer the case for the extended charged Higgs
mass range considered here, and thus the ratio between these two
observables can be different.  

\begin{widetext}

\begin{figure}
	\hspace*{\fill}
	\subfloat[]{
		\includegraphics[width= .45 \textwidth]{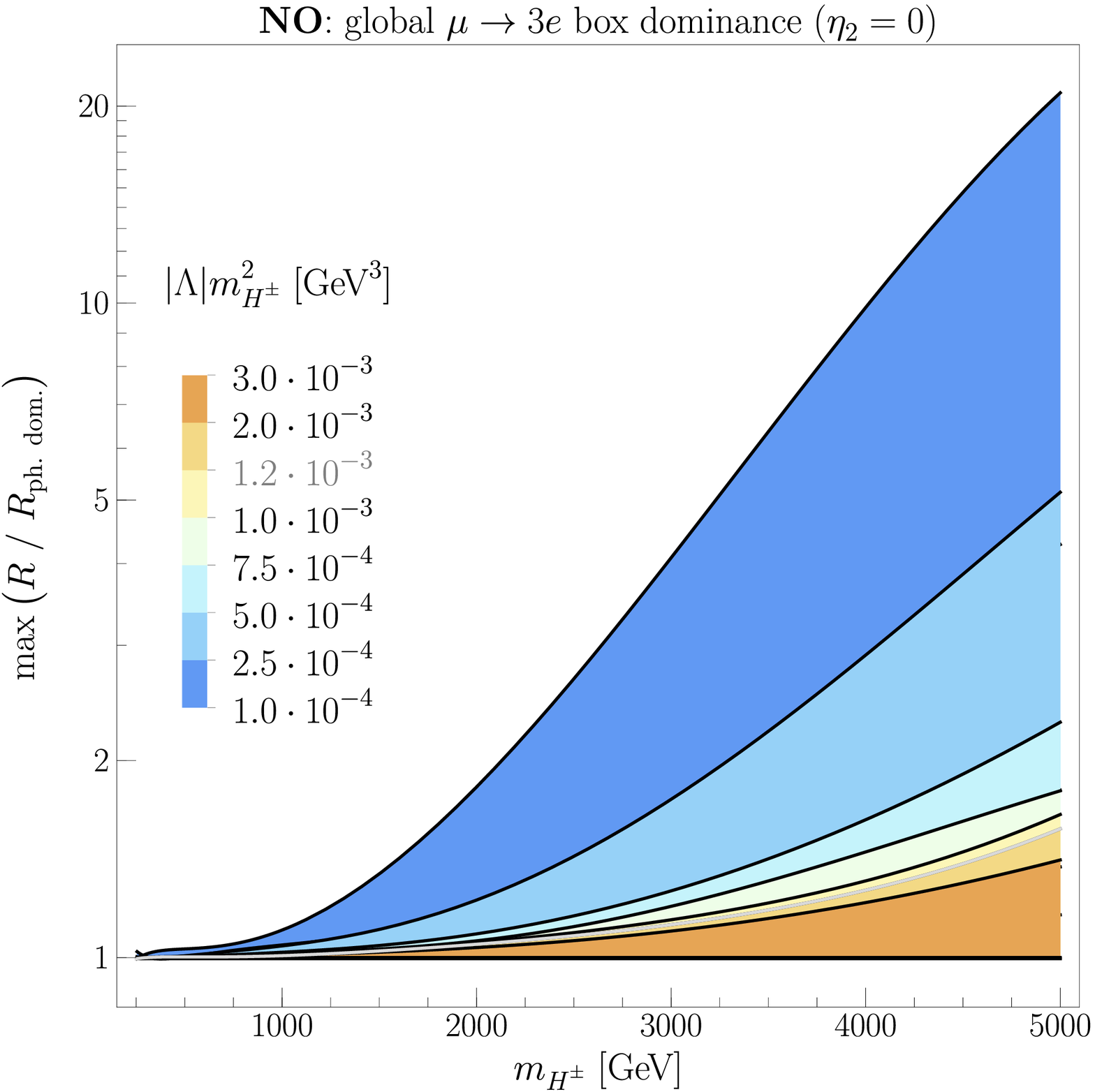} }
	\hfill
	\subfloat[]{
		\includegraphics[width= .45 \textwidth]{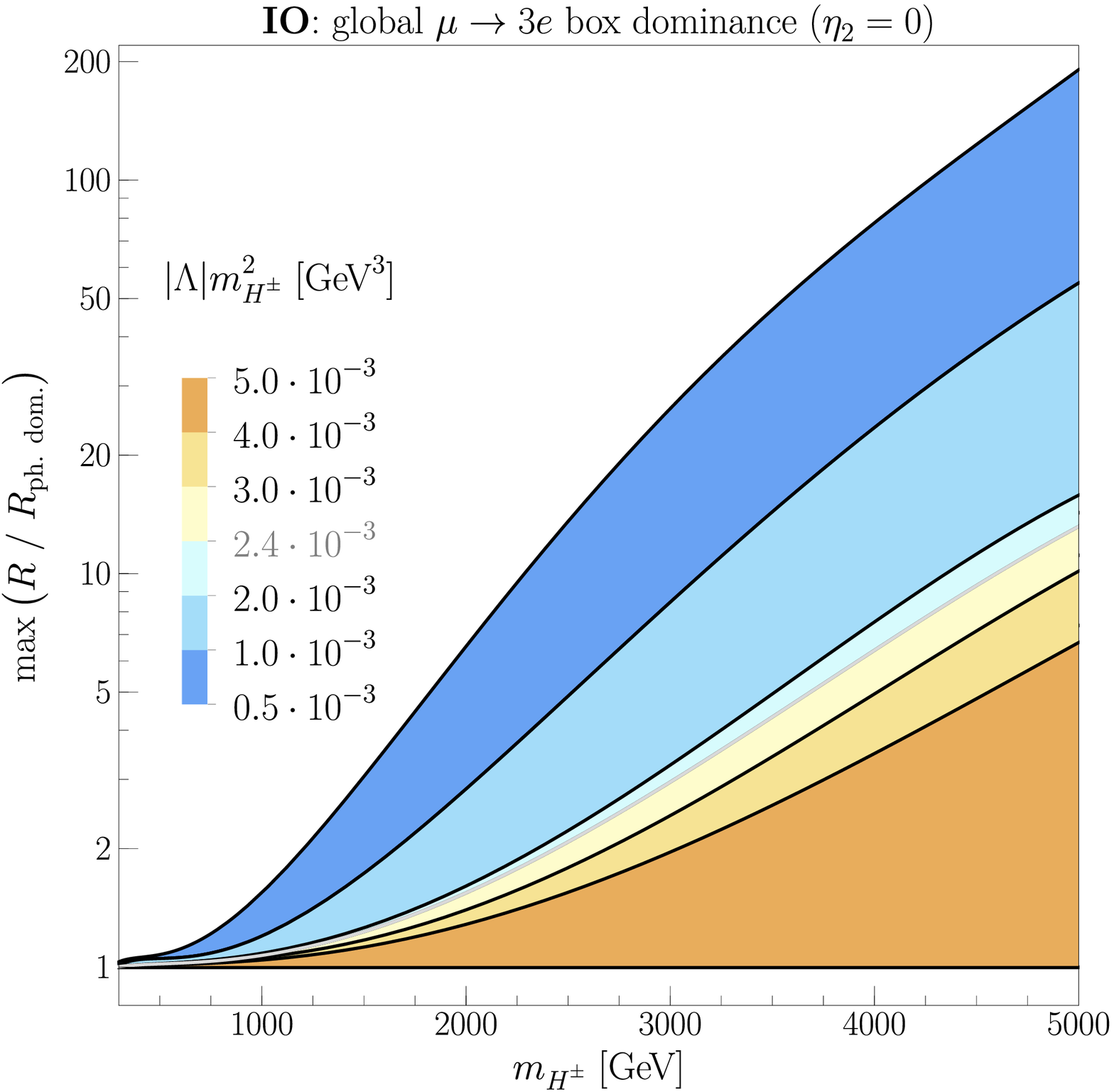} }
	\hspace*{\fill}
	\caption{
	Maximal currently allowed deviation from photon dominance as a function of \chargedMass{} in the typical parameter region. 
	It is plotted as a maximal currently allowed
	ratio of $R\equiv \br(\mueee{})/ \br(\mueg{})$, normalized to the ratio \photondominance{}, calculated from Eq.~\eqref{eq:photon dominance}, i.e. when the photon dominance is assumed. 
	Colors show photon factor values for which these deviations can occur.
	}\label{fig:global-meee}
\end{figure}

\end{widetext}

We plot the possible ratios of \mueee{}/\mueg{} for fixed \br
(\mueg{}) at current limits (normalized to the photon dominance value) for
different photon factor values in figure~\ref{fig:global-meee}. 
The colors for photon factor values are consistent with the coloring
scheme of figure~6 of Ref.~\cite{Dudenas:2022von} and correspond to the same regions in \plane{} of that reference.
The very highest values (darkest blue) come from a region close to the special region, in which $\flavorY2{\mu}\approx 0$. 
This can be easily understood by recalling that photonic amplitudes are proportional to $\flavorY2{\mu}\flavorY2{e}$, while boxes $\propto\flavorY2{\mu} \big(\flavorY2{e}\big)^3$, thus the photonic contributions are suppressed relative to the box ones. 
Getting further away from this region, we get lower and lower box enhancement. 
The regions close to $\flavorY2{e}\approx 0$ have box contributions suppressed, thus correspond to photon dominance. 
They also correspond to a smaller photon factor value (of dark blue). 
However, they are not visible in figure~\ref{fig:global-meee}
as they are hidden behind the largest parameter space with high photon factor value (orange). 

In general,
the deviations from photon dominance are caused by large Yukawa
couplings which in turn are correlated with small $\Lambda$, see
Eq.~\eqref{eq:Y2 mass eigenstates+}. In figure~\ref{fig:global-meee}, therefore, small
photon factor and/or large charged Higgs mass correlates with strong
deviations from photon dominance, as can be seen from Eq.~\eqref{eq:box/penguin}. For example, 
outside the darkest blue region with photon factor less than $10^{-4}$,
deviations from photon dominance by a factor of two can be reached if the charged Higgs mass is
above 3 TeV.

\subsection{Yukawa ``fingerprint'' of the model ($\majoPhase{} = 0$)}

So far we have employed the advantageous parameterization of Yukawa
couplings in terms of the \plane. Now we finally show how the model
constrains affect the actual values of the Yukawa couplings \flavorY2{}. Since
these are the fundamental Lagrangian parameters, constraints on them
are of interest since they might shed light on possible patterns and
fundamental origins of the Yukawa couplings.

More specifically, we look at the minimum vs. maximum values of the
components of the three-vector \flavorY2{} as a two-dimensional
scatter plot. In this way we show which overall magnitudes of Yukawa
couplings are possible, and also whether the Yukawa couplings
can/should involve strong hierarchies or not.

\begin{widetext}

\begin{figure}[]
	\hspace*{\fill}
	\subfloat[]{
		\includegraphics[width=.45 \textwidth]{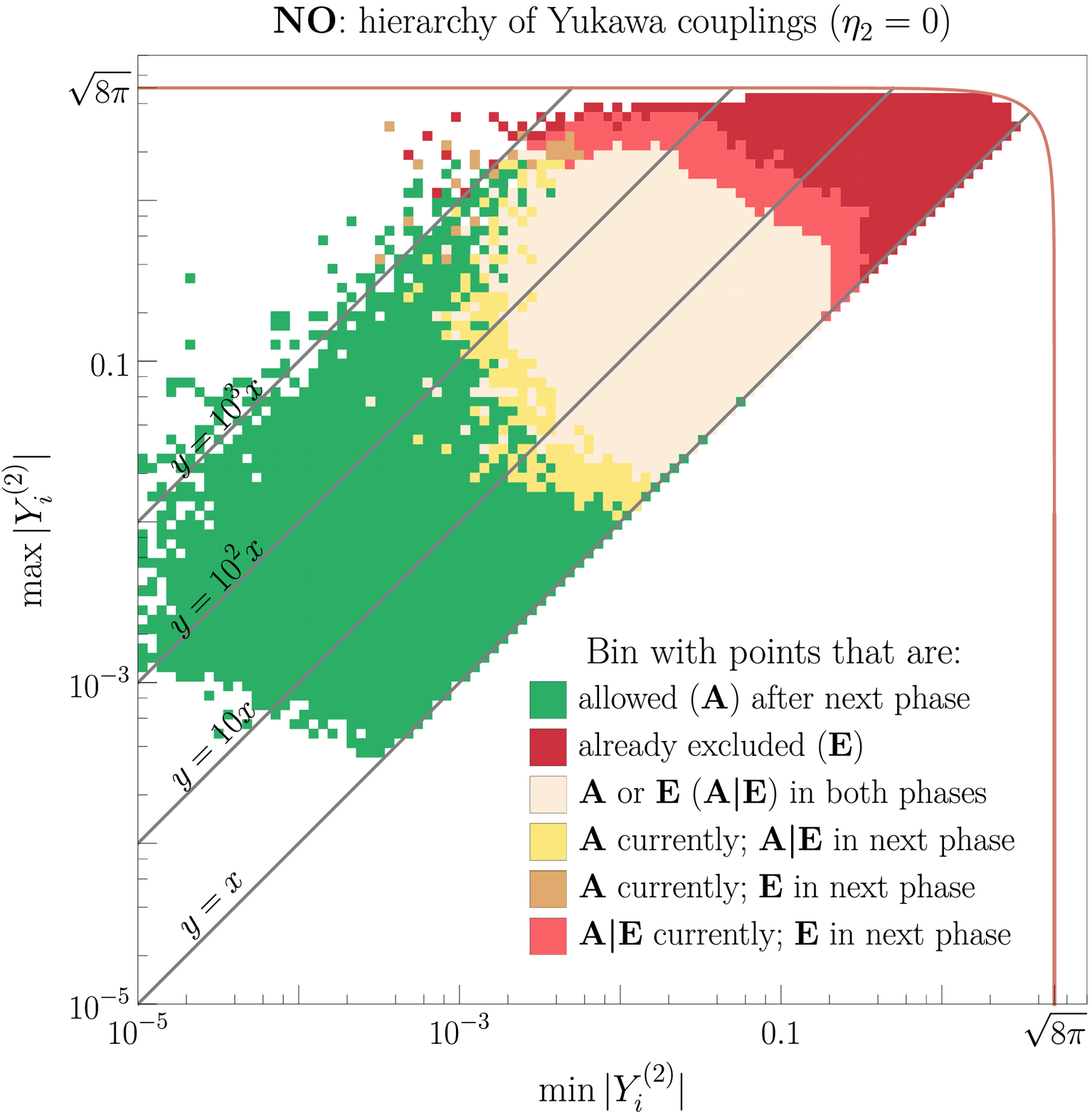} }
	\hfill
	\subfloat[]{
		\includegraphics[width=.45 \textwidth]{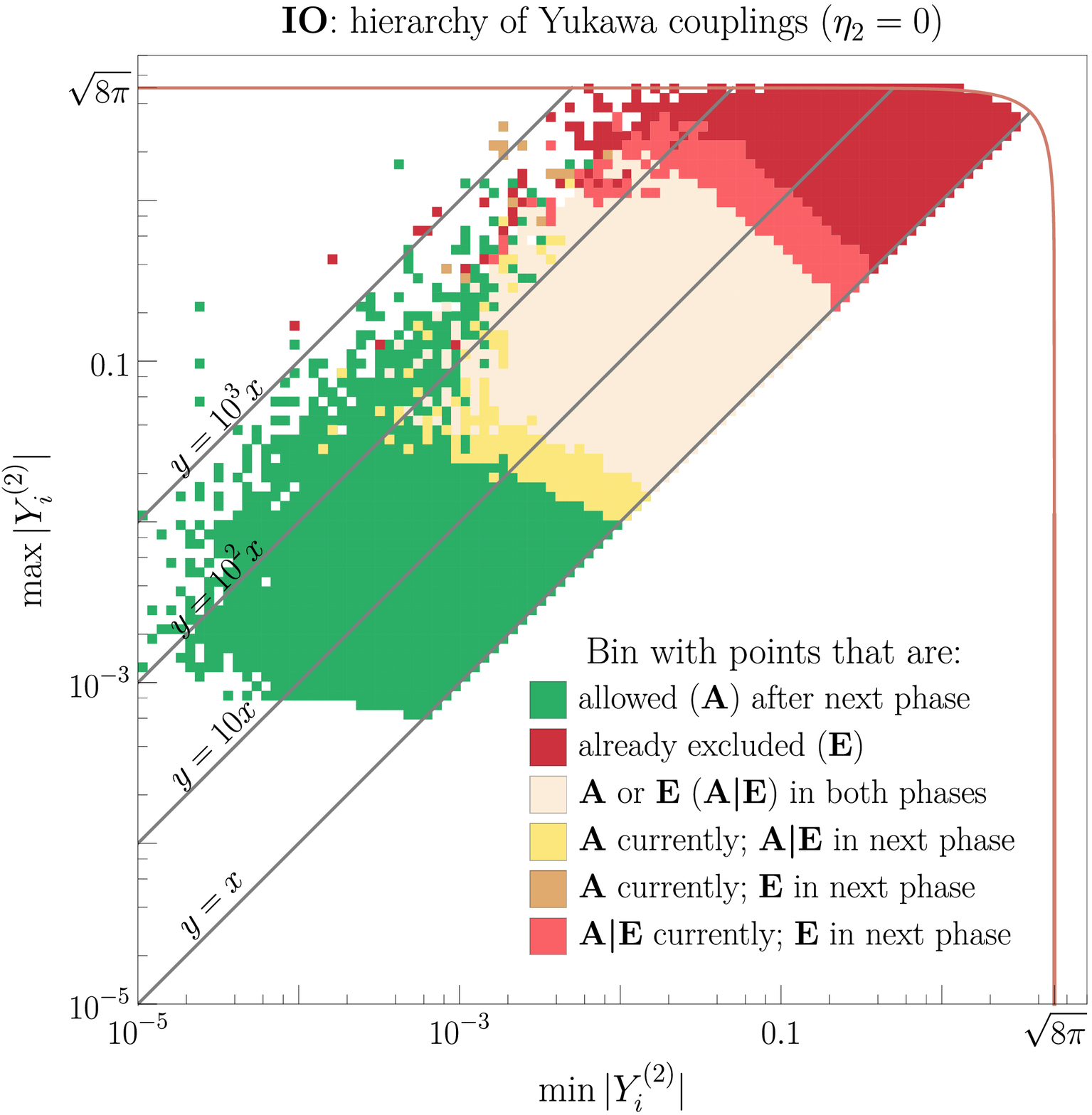} }
	\hspace*{\fill}
	\caption{Constrains on Yukawa couplings from \gls{clfv}. 
		The Yukawa couplings, which are always allowed, sometimes allowed (depending on other variables), and always excluded by both current and next phase experiments are colored by green, yellow and red respectively. 
		The regions, which change from ``allowed'' to ``ambiguous'' and from ``ambiguous'' to ``excluded'' after next phase experiments are indicated by more intensive yellow and red colors. 			
	}\label{fig:minmax-zeroeta}
\end{figure}

\end{widetext}

The resulting plot is given in figure~\ref{fig:minmax-zeroeta}. 
Every point in the plot corresponds to a bin of Yukawa values, while
other parameters are scanned
over, see Eq.~(\ref{eq:free params}).
Different colors correspond to allowance/exclusions of current/next phases of
experiments, which could be captured by a scan in the parameter space.  
The diagonal lines separate the orders of magnitude of hierarchies
between the Yukawa couplings.  
For instance, the very lowest diagonal line corresponds to the case, in which all the components of \flavorY2{} are equal, while the leftmost corresponds to the $O(10^3)$ ratio between the minimum and maximum value of \flavorY2{}. 
The special points with $\flavorY2{\mu,e}=0$ discussed in previous
subsections would be located to the left of the visible plot range in
figure \ref{fig:minmax-zeroeta}.

The dark red region in figure~\ref{fig:minmax-zeroeta} is
excluded already by current phase experiments. In this completely excluded
region the Yukawa couplings take rather large values; everywhere in
the dark red region at least one Yukawa coupling is above $0.1$.

The dark green region is experimentally allowed and will remain
allowed even if next phase experiments see no signal. As expected this
region contains smaller Yukawa couplings; though non-hierarchical
Yukawa couplings are possible up to values of $10^{-2}$ and
hierarchical Yukawa couplings are possible where the minimum/maximum
Yukawa values are around $10^{-3}\div 1$. 
The yellow region, and the regions with different lighter shadings of
yellow and red, are \emph{ambiguous}. Each bin in these regions
contains points which are allowed now by current phase experiments and
points which can be excluded by next phase experiments, as explained in
the legend of the figure. This region is thus the region which will be
scrutinized by next phase experiments, and part of this region can be
excluded. 
In the yellow region there
exist parameter points where all Yukawa couplings are ${\cal O}(0.1)$
as well as hierarchical points where the minimum/maximum Yukawa
coupling values are around $10^{-3}\div 1$. 

\section{Conclusions}\label{sec:conclusions}
We completed the \gls{clfv} study of the \gls{gnm} by investigating its full parameter space. 
The enlarged parameter space, compared to the previous
study~\cite{Dudenas:2022von}, is summarized in
table~\ref{tab:extension of params}. The \gls{gnm} remains a favorable
explanation of neutrino masses in the parameter region of tiny seesaw
scale, i.e.\ a seesaw scale below the electroweak scale. In addition, this leads to a prediction of two- and three-body decay \gls{clfv} processes and provides reasonable restrictions on the scalar and leptonic parameters.

Just as in the previous study, the weakest and the most conservative
(``absolute'') bounds on the photon factor are provided by $\tau{}$ two- and three-body decays in special regions in which one of the Yukawa couplings vanishes $\flavorY{2}{e,\mu}\approx 0$.
The impact of the Majorana phase \majoPhase{} can be seen in figure~\ref{fig:new-two-body-absolute-limits}. 
It does not lead to a drastic change in
the two-body decay bounds, i.e. the bounds  stay within an order of
magnitude as a function of \majoPhase{}. 
Also, the absolute bound for $\majoPhase{}=0$ turns out to be very close to the lowest possible value.
This means that the result for the absolute bound obtained in
\cite{Dudenas:2022von} remains the absolute lower bound for general \majoPhase{}  to a very good approximation. 

In the parameter regions with \chargedMass{} above the TeV scale,
larger Yukawa couplings are possible, see figure \ref{fig:three-body-bounds}. 
As a result, box diagram contributions governed by four powers of Yukawa couplings dominate the three-body decay observables, in contrast to the photonic dominance for the low-mass region.
For high \chargedMass{} and  in the special regions ($\flavorY2{e,\mu}\approx 0$), perturbative unitarity constraints for Yukawa couplings, Eq.~\eqref{eq:pert unitarity}, can become more relevant than \gls{clfv} ones.

Figure~\ref{fig:factor-bounds} summarizes the main constraints
for $\majoPhase{}=0$, which are the restrictions in the \factorplane{} from both two and three-body decays.
The figure shows that the parameter space is constrained by an
interplay between box-dominated three-body decays, two-body decays, and perturbative unitarity. 
Thus we update the absolute and typical bounds in Eq.~\eqref{eq:absolute} and in Eq.~(\ref{eq:typical}). 

Outside of the special regions where specific Yukawa couplings vanish
and where $\tau$ decays are important, $\mueg$ still restricts the
majority of the parameter space due to stronger experimental
bounds. Also, complementary to this two-body decay, $\mueee$ will
become competitive already in the next phase (Mu3e-I) and will become
even more restrictive at Mu3e-II. In addition to the purely experimental arguments mentioned above, the latter observable is affected by the box contributions that are enhanced for larger values of \chargedMass{}.
This allows for significant deviations from photon dominance, which are shown in figure~\ref{fig:global-meee} for Mu3e-I. 

We recall that the scoto-seesaw model and the \gls{gnm} in the tiny seesaw scale have the same predictions for \gls{clfv} and thus all our results are directly applicable for a scoto-seesaw model, too.
One should expect the same qualitative behavior for the scotogenic model in these parameter regions, as argued in~\cite{Dudenas:2022von}.
Thus by studying the \gls{gnm}, we also complement the previous studies of \gls{clfv} of scoto-seesaw and scotogenic models by including the tiny sterile neutrino mass region. 

\acknowledgments
U.Kh. was supported by the Deutscher Akademischer Austauschdienst (DAAD) under Research Grants \textemdash{} Doctoral Programmes in Germany, 2019/20 (57440921)
and by the DFG under grant number STO 876/7-1.
W.K. was supported by the National Science Centre (Poland) under the research grant 2020/\allowbreak38/\allowbreak E/\allowbreak ST2/\allowbreak 00126.
V.D. and T.G. thank the Lithuanian Academy of Sciences for
their support via project DaFi2021.

\bibliographystyle{apsrev4-2}
\bibliography{bibliography}

\end{document}